\documentclass[aps,reprint,showpacs]{revtex4-1}
\usepackage{xcolor}
\usepackage{xfrac}
\usepackage{amsmath}
\usepackage{graphicx}
\usepackage[yyyymmdd,hhmmss]{datetime}
\usepackage{lmodern}
\usepackage[T1]{fontenc}
\usepackage[normalem]{ulem}

\newcommand{\uvec}[1]{\boldsymbol{\hat{\textbf{#1}}}}

\begin{document}

\title{Rayleigh waves, surface disorder, and phonon localization in nanostructures}

\author{L. N. Maurer}
\email{lnmaurer@wisc.edu}
%\affiliation{University of Wisconsin-Madison, Madison, Wisconsin 53706, USA}

\author{S. Mei}

%\affiliation{University of Wisconsin-Madison, Madison, Wisconsin 53706, USA}

\author{I. Knezevic}
\email{iknezevic@wisc.edu}
\affiliation{University of Wisconsin-Madison, Madison, Wisconsin 53706, USA}
\date{\today}

\begin{abstract}
%This version was compiled on \today\ at \currenttime.\\

We introduce a technique to calculate thermal conductivity in disordered nanostructures: a finite-difference time-domain (FDTD) solution of the elastic wave equation combined with the Green-Kubo formula. The technique captures phonon wave behavior and scales well to nanostructures that are too large or too surface disordered to simulate with many other techniques. We investigate the role of Rayleigh waves and surface disorder on thermal transport by studying graphenelike nanoribbons with free edges (allowing Rayleigh waves) and fixed edges (prohibiting Rayleigh waves). We find that free edges result in a significantly lower thermal conductivity than fixed ones. Free edges both introduce Rayleigh waves and cause all low-frequency modes (bulk and surface) to become more localized. Increasing surface disorder on free edges draws energy away from the center of the ribbon and toward the disordered edges, where it gets trapped in localized surface modes.  These effects are not seen in ribbons with fixed boundary conditions and illustrate the importance of phonon surface modes in nanostructures.
\end{abstract}


\maketitle

\section{Introduction}

Surface disorder can disrupt phonon thermal transport throughout a nanostructure resulting in drastically reduced thermal conductivity \cite{Yang_Nat_08, Yang_NLet_12, Knezevic_APL_15}. However, in spite of considerable recent interest in phonon transport at the nanoscale \cite{Cahill_APR_14}, the effects of disordered nanostructure surfaces are poorly understood. In particular, surface modes and their effect on transport have received limited attention.

\begin{figure}
 \centering
  \includegraphics[width=3 in]{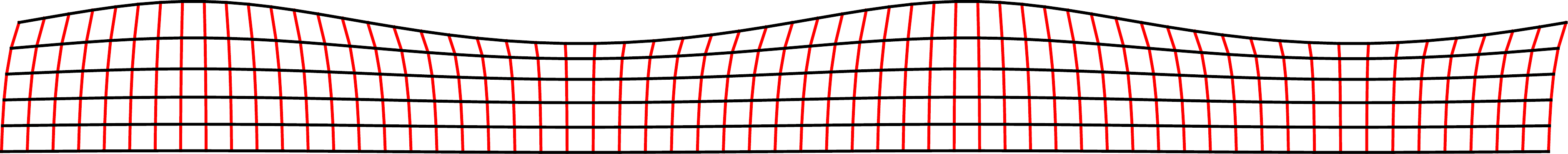}
 \caption{Illustration of a Rayleigh wave propagating along the free top surface. The wave amplitude decays exponentially with increasing distance from the surface.}
 \label{fig:Rayleigh_wave}
\end{figure}

Rayleigh waves (Fig. \ref{fig:Rayleigh_wave}) are a simple type of surface mode that occur only at free surfaces \cite{Rayleigh_PLMS_1885, Landau_Elasticity, Srivastava_Book, Wallis_PSS_74, Wallis_SS_94} and provide a good starting point for understanding surface modes in general. It has long been known that bulk waves (longitudinal and transverse) can be converted to Rayleigh waves at rough \cite{Falkovskii_JTEP_73, Maznev_PRB_15} or otherwise disordered \cite{Klemens_PRL_70, Sakuma_PRB_73, Nakayama_PRB_85} surfaces, but not at smooth ones \cite{Wolfe_Book, Landau_Elasticity}. So, early work on Rayleigh wave scattering focused on their decay into bulk waves in the presence of disorder \cite{Klemens_PRL_70, Sakuma_PRB_73, Falkovskii_JTEP_73, Maradudin_PoP_87, Maradudin_NPD_87}, but did not address phonon transport. Nakayama \cite{Nakayama_PRB_85} identified Rayleigh-to-bulk mode conversion as a cause of diffuse phonon--surface scattering, which is important to many phonon transport models \cite{Ziman_PRS_53, Ziman_PRS_55, Ziman_Book}. Maznev \cite{Maznev_PRB_15} investigated elastic wave scattering from a nearly smooth surface using a Green's-function technique and found that most energy from a normally incident longitudinal wave can be converted into Rayleigh waves, but conversion at other angles was not addressed. Kang and Estreicher \cite{Estreicher_PRB_14} used molecular dynamics (MD) simulation to show that mode conversion between bulk modes and surface modes can lead to ``phonon trapping,'' which can greatly reduce thermal conductivity. Localized modes associated with disorder play an important role in reducing thermal conductivity \cite{Estreicher_JAP_15}.

Elastic continuum materials provide excellent model systems for studying surfaces and their effects on transport and localization. Elastic materials are a simpler, long-wave-length limit of atomic materials \cite{Ashcroft_Mermin, Srivastava_Book}, and elastic materials support longitudinal acoustic, transverse acoustic, Rayleigh \cite{Landau_Elasticity, Srivastava_Book, Wallis_PSS_74, Wallis_SS_94}, and structure-dependent modes (such as torsional modes for a wire \cite{Tamura_PSSc_04}). The study of elastic continuum materials can provide important insights into phonon--surface scattering in atomic materials \cite{Srivastava_Book, Cross_PRL_01, Gu_JoPCM_04, Jang_JAP_12, Lee_APL_12, Maznev_PRB_15}, especially in nanostructures that are too large to treat directly using atomistic techniques, yet too small to be considered bulk \cite{Roche_PRB_10,Cuniberti_PRB_10,Keblinski_APL_10,Cahill_APR_14}.

In this paper, we examine the effects of surface disorder and Rayleigh waves on thermal conductivity in nanostructures. We model two-dimensional (2D) graphenelike elastic nanoribbons with random, rough edges. By switching between fixed and free boundary conditions (BCs) we can selectively ``turn off'' Rayleigh waves while keeping the surface disorder. We solve the elastic wave equations with the finite-difference time-domain (FDTD) technique, which is computationally efficient and not limited to weak roughness. We couple the FDTD simulation with the Green-Kubo formula to calculate the thermal conductivity. This combination of FDTD and Green-Kubo has not been used before and has similarities to an equilibrium molecular dynamics (MD) simulation \cite{Hoover_PRB_86, Yip_JNucM_98} but scales better than MD as the system size increases. This combined technique enables us to simulate large nanostructures with pronounced roughness. We also perform a vibrational eigenmode analysis \cite{Feldman_PRB_93, Li_NLet_10, Russ_EPL_04} to further investigate the connection between boundary conditions and localization. Not surprisingly, increasing surface roughness decreases the thermal conductivity and increases localization. However, the inclusion of free surfaces (and the associated Rayleigh waves) further lowers the thermal conductivity and causes low-frequency bulk modes to become more localized. As the roughness is increased, energy is drawn from the center of the ribbon to the edges, which are the most disordered parts of the system and are where the energy is least likely to propagate. We also investigate the effects of mode conversion at surfaces, the limitations of Casimir's phonon-surface scattering model \cite{Casimir_38, Ziman_PRS_55}, and the limitations of models for phonon--surface scattering that rely on a single scalar wave and do not include mode conversion or Rayleigh waves.

In Sec. \ref{sec:theory}, we review the theory of elastic waves. We include subsections on the elastic wave  and scalar wave equations (\ref{sec:equations}), the utility of the continuum limit in 2D materials (\ref{sec:continuum_limit}), physically relevant boundary conditions (\ref{sec:BCs}), and an overview of Rayleigh waves (\ref{sec:Rayleigh}).  In Sec. \ref{sec:method}, we introduce the FDTD simulation technique, the structures we simulate (\ref{sec:FDTD}), and explain how the method can be used with the Green-Kubo formula to compute thermal conductivity (\ref{sec:kappa}). Based on the FDTD simulation, we illustrate how the interplay of the boundary conditions (free or fixed) with surface roughness affects the mode conversion channels and energy trapping near the surface (Sec. \ref{sec:mode_conversion}). We investigate the effects of boundary conditions on localization in Sec. \ref{sec:VEA} using vibrational eigenmode analysis. In Sec. \ref{sec:results}, we report the calculated thermal conductivities for elastic and scalar wave equations with different types of boundary conditions, and we interpret these results in Sec. \ref{sec:Casimir} and Sec. \ref{sec:length}. In Sec. \ref{sec:energy_localization}, we demonstrate that rough surfaces siphon energy from inside the nanostructure and trap it near the surface. In Sec. \ref{sec:implications} we discuss the implications of Rayleigh waves and mode conversion on common phonon--surface scattering models. Finally, we conclude with Sec. \ref{sec:discussion}.

\section{Theoretical framework}\label{sec:theory}

\subsection{Elastic and scalar wave equations}\label{sec:equations}

We consider isotropic, homogeneous, linearly elastic, continuum materials \cite{Landau_Elasticity, Brekhovskikh_Book}. $\mathbf{u}\left(\mathbf{r},t\right)$ is the displacement of an infinitesimal element of the material. The strain tensor

\begin{subequations}
\begin{equation}\label{equ:strain}
  \sigma_{ij} = \frac{1}{2}\left(\partial_j u_i + \partial_i u_j\right)
\end{equation}

\noindent
and the stress tensor $\tau_{ij}$ are related by the continuum generalization of Hooke's law

\begin{equation}\label{equ:Hooke}
  \tau_{ij} = \lambda \sigma_{ll} \delta_{ij} + 2\mu\sigma_{ij},
\end{equation}
\end{subequations}

\noindent
where $\lambda$ and $\mu$ are the Lam\'e parameters, which are material dependent. $\mu$ is also known as the shear modulus and $\mu>0$. Thermodynamic stability requires $\lambda > -\frac{2}{3} \mu$ \cite{Landau_Elasticity}, but $\lambda > 0$ for most materials \cite{Maradudin_NPD_87}.

Applying Newton's second law to the stress--strain relations yields
\begin{align}\label{equ:elastic_wave}
    \begin{split}
        \rho{ \ddot{\mathbf{u}}} &= \mathbf{\nabla} \cdot \mathbf{\tau} \\
        &= ( \lambda + 2\mu )\nabla(\nabla \cdot \mathbf{u}) - \mu\nabla \times (\nabla \times \mathbf{u}),
    \end{split}
\end{align}
\noindent
where $\rho$ is the density. We will refer to Eq. (\ref{equ:elastic_wave}) as the \textit{elastic wave equation}, and we will refer to a (vector field) solution of the elastic wave equation as an \textit{elastic wave}.

The power flux density is \cite{Wolfe_Book, Snieder_02}
\begin{equation}\label{eq:J}
  \mathbf{J} = - \mathbf{\tau}\cdot\dot{\mathbf{u}}.
\end{equation}
$\mathbf{J}$ is the elastic wave equivalent of the Poynting vector for electromagnetic waves. (In the context of thermal transport, $\mathbf{J}$ is commonly referred to as the heat current density.)

We will also consider wave equations of the form $\nabla^2 \phi - \frac{1}{c}\ddot{\phi} = 0$, which we will refer to as the \textit{scalar wave equation}, and we will refer to a solution of the scalar wave equation as a \textit{scalar wave}.

By Helmholtz's theorem, $\mathbf{u}$ can be written in terms of a scalar potential $\phi$ and a vector potential $\mathbf{\Psi}$ \cite{Landau_Elasticity, Brekhovskikh_Book}

\begin{equation}\label{equ:decompostition}
 \mathbf{u} = \nabla\phi + \nabla\times\mathbf{\Psi},
\end{equation}

\noindent
which allows the elastic wave equation (\ref{equ:elastic_wave}) to be decomposed into two wave equations, one for longitudinal and one for transverse waves \cite{Landau_Elasticity, Brekhovskikh_Book}:
\begin{subequations}\label{eqn:scalar_waves}
  \begin{align}
        \nabla^2 \phi - \frac{1}{c_l^2}\ddot{\phi} & = 0, \\
        \nabla^2 \mathbf{\Psi} - \frac{1}{c_t^2}\ddot{\mathbf{\Psi}} & = 0,
  \end{align}
\end{subequations}

\noindent
where

\begin{subequations}\label{eqn:group_velocities}
  \begin{align}
        c_l &= \sqrt{\frac{\lambda + 2\mu}{\rho}}, \\
        c_t &= \sqrt{\frac{\mu}{\rho}},
  \end{align}
\end{subequations}

\noindent are the longitudinal and transverse wave speeds, respectively.

Consider a three-dimensional (3D) slab in the $xy$ plane, effectively a quasi-2D system. For transverse waves in the $xy$ plane, $\mathbf{\Psi}$ will always be normal to the plane, i.e., $\mathbf{\Psi}_x = \mathbf{\Psi}_y = 0$ and $\psi = \mathbf{\Psi}_z$ is a scalar field. Thus, the elastic wave equation (\ref{equ:elastic_wave}) for longitudinal and transverse waves in a quasi-2D slab can be decomposed into two scalar wave equations, which are independent in an infinite medium and whose solutions are scalar fields $\phi$ and $\psi$, respectively.

To study 2D materials rigorously, we take the zero-thickness limit of the quasi-2D, free-standing-slab model. The result is an ``elastic plane'', which has the same in-plane stress--strain relations as the quasi-2D model, except that $\lambda$ is suitably modified. To transform the quasi-2D model into a 2D,  elastic-plane model, we make the following substitutions in Eq. (\ref{equ:Hooke}) \cite{Landau_Elasticity}:

\begin{subequations}\label{eqn:2D_substitutions}
  \begin{align}
        \lambda & \rightarrow \frac{2\lambda^\prime \mu^\prime}{\lambda^\prime + 2 \mu^\prime} , \\
        \mu     & \rightarrow \mu^\prime ,
  \end{align}
\end{subequations}

\noindent where the primed quantities are the Lam\'e parameters for the 3D material that is being formed into a thin plate, and the unprimed quantities can be thought of as 2D Lam\'e parameters. Materials like single-layer graphene are inherently 2D, so the unprimed 2D Lam\'e parameters are generally reported in the literature (in which case the substitutions in Eq. (\ref{eqn:2D_substitutions}) are obviously unnecessary).

We do not model any out-of-plane modes in this work, but we note that an elastic plane has a single out-of-plane mode. The out-of-plane mode is described a scalar wave equation of the form $\nabla^4 \Phi \propto \ddot{\Phi}$, which results in a quadratic dispersion relation \cite{Landau_Elasticity}. This quadratic dispersion relation is seen in the long-wave-length limit of graphene ZA modes \cite{Kresse_PRB_03}. The out-of-plane mode is decoupled from the in-plane modes \cite{Landau_Elasticity}, so the out-of-plane mode will not display the mode conversion or surface waves investigated in this work.

As the elastic wave equation in the bulk can be decomposed into two simpler scalar wave equations, scalar waves are used more often than elastic waves to model phonon--surface scattering  \cite{Lambert_PRB_99, Cross_PRB_01, Lifshitz_PRB_01, Yung_PRB_01, Vasilopoulos_PRB_01, Geller_PRB_04, Chang-Long_CPL_06, KChen_PLA_06, KChen_JoPD_07, Xie_JoPD_07, Zhao_JoPD_07, JGong_PRB_09}. However, the decomposed waves are only independent of each other in an infinite medium. The boundary conditions at surfaces couple the waves together \cite{Brekhovskikh_Book}. Without the proper boundary conditions, the wave equations erroneously remain independent, which means they do not capture mode conversion (Sec. \ref{sec:mode_conversion}), Rayleigh waves (Sec. \ref{sec:Rayleigh}), or many structure-dependent waves \cite{Tamura_PRB_05}.

\subsection{The continuum limit of 2D materials}\label{sec:continuum_limit}

The continuum limit is the regime where long-wave-length acoustic phonons in atomic materials behave like elastic waves in continuum materials, i.e., $\omega \approx c |\vec{k}|$, where $\omega$ is the angular frequency, $\vec k$ is the wave vector, and $c$ is the wave speed. (The elastic waves and elastic wave equation would have to be generalized to allow for anisotropy \cite{Ashcroft_Mermin, Srivastava_Book}).

\begin{figure*}
 \centering
 \includegraphics[width=\textwidth]{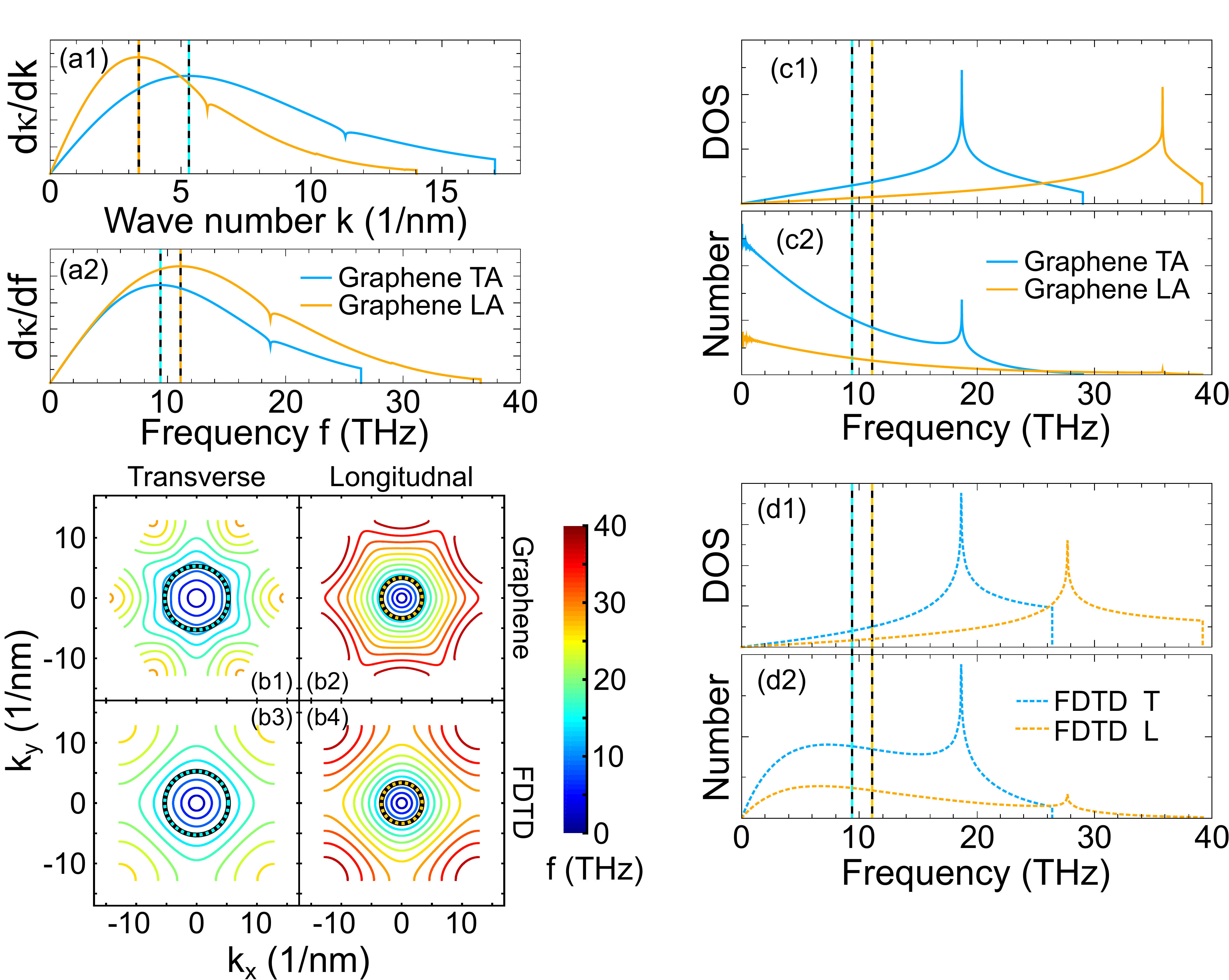}
 \caption{Phonon dispersions and thermal transport in graphene and in our graphenelike model system simulated via FDTD. All temperature-dependent plots are at 300 K. (a1) \& (a2) The contribution to thermal conductivity $\kappa$ from phonons with wave numbers around $k$ (a1) or frequency around $f$ (a2). Thermal conductivity is calculated from the RTA and standard umklapp and isotope scattering rates, with scattering lifetime having an upper bound due to the finite size of the sample. The maximum contributions for transverse acoustic (TA) and longitudinal acoustic (LA) phonons occur at wave numbers 3.37 and 5.28 $\text{nm}^{-1}$ (a1) (wave lengths of 1.86 and 1.19 nm) and  frequencies 9.33 and 11.0 THz (a2), respectively. Dashed black-blue and black-orange lines in all panels indicate the phonon wave number or frequency that contribute the most to thermal conductivity from TA and LA modes, respectively.  (b1)--(b4) Contour plots of the phonon dispersion relations for graphene TA (b1), graphene LA (b2) \cite{Knezevic_JAP_14}, FDTD transverse (b3), and FDTD longitudinal (b4) modes [Eq. (\ref{equ:dispersion})]. Dashed black-blue and black-orange circles are the constant-wave-number curves for the peak wave numbers in (a1). (c1) The density of states (DOS) for TA (blue) and LA (orange) modes in graphene, calculated based on full phonon dispersions  \cite{Knezevic_JAP_14}. (c2) The number of phonons per unit frequency for graphene, calculated based on the full-dispersion DOS from (c1) and Bose-Einstein statistics. A large fraction of the graphene phonon population falls in the continuum limit. (d1) The DOS for transverse (blue dashed) and longitudinal (orange dashed) modes in our graphenelike model system. (d2) The number of phonons per unit frequency in our graphenelike model system based on the DOS from (d1). The system is classical, so the phonon number per unit frequency is calculated with Maxwell-Boltzmann statistics and tends to zero at low frequencies. However, a large fraction of the phonon population is still in the continuum limit, and the population is sizeable at the frequencies of maximum contribution to thermal conductivity.} \label{fig:NormalizedNumbers}
\end{figure*}

The continuum limit is much more important for 2D materials than for their 3D counterparts. Namely, the number of phonons per unit angular frequency $\omega$ is $N\left(\omega\right)=g\left(\omega\right)n_{\mathrm{BE}}\left(\omega\right)$, where $n_{\mathrm{BE}}$ is the Bose-Einstein distribution and $g$ is the density of states. When $\omega$ is small,  $n_{\mathrm{BE}}\left(\omega\right) \approx k_b T / \hbar \omega$ (the equipartition approximation). In 3D, the density of states $g_{3D}\left(\omega\right) \propto \omega^2$, so $\lim_{\omega\to 0} N_{3D}\left(\omega\right) = 0$, i.e., there are relatively few phonons in the continuum limit in 3D \cite{Ashcroft_Mermin}. In contrast, the continuum limit of in-plane 2D modes yields the density of states $g_{2D}\left(\omega\right) \propto \omega$, so $\lim_{\omega\to 0} N_{2D}\left(\omega\right)$ is nonzero \cite{Mingo_PRB_10,Knezevic_JAP_14}. The result is clear in graphene at 300 K [Fig. \ref{fig:NormalizedNumbers}(c2)]: a large fraction of the phonons are in the continuum limit.  Phonons in the continuum limit also have the highest group velocity of any phonons in the material, along with relatively low scattering rates owing to their long wave lengths \cite{Mingo_PRB_05, Broido_PRB_10}. For these reasons -- abundance, high group velocity, and low scattering rates -- in-plane phonons in the continuum limit are of great significance for the thermal conductivity in 2D materials \cite{Balandin_APL_09, Balandin_PRB_09}.

The importance of long-wave-length phonons for thermal transport in 2D materials can be grasped by analyzing thermal conductivity in graphene in the relaxation time approximation (RTA). The thermal conductivity $\kappa$ along a certain direction, as calculated within the RTA, is given by \cite{Carruthers_RMP_61}

\begin{equation}\label{equ:RTA}
 \kappa \propto\int k\,dk\int d\theta\, \sum_j \tau_{j,\vec{k}} C_{j,\vec{k}} {v}^2_{j,\vec{k}}\cos^2{\theta},
\end{equation}

\noindent
where the sum goes over the different branches $j$, $\tau_{j,\vec{k}}$ is the relaxation time, $\vec{v}_{j,\vec{k}}$ is the group velocity and $\theta$ is its angle with respect to the direction of heat flow, and $C_{j,\vec{k}}$ is the heat capacity per mode. The relaxation times in graphene for umklapp phonon--phonon and isotope scattering are $\tau_{j,\vec{k}} \propto \omega^{-2}$ \cite{Balandin_APL_09, Knezevic_JAP_14}. In the $k\to 0$ limit, $\omega \propto k$, thus $\tau_{j,\vec{k}} \propto k^{-2}$. As  $v_{j,\vec{k}} \to c_j$ and $C_{j,\vec{k}} \to k_B$ ($k_B$ is the Boltzmann constant), the divergence in $\tau_{j,\vec{k}}$ causes the integrand of the wave-number integral [Eq. (\ref{equ:RTA})] to diverge at small $k$, and $\kappa$ becomes infinite largely due to the contribution of long-wave-length phonons \cite{Balandin_APL_09}.

In reality, $\tau_{j,\vec{k}}$ has a behavior closer to $\omega^{-1}$ at small frequencies \cite{Marzari_NLet_12}, which will fix the above divergence while further emphasizing the important contribution from long-wave-length modes. Moreover, no experimental sample has an infinite size, so the lifetime of a low-energy, long-wave-length phonon is ultimately never infinite; rather, it is limited by scattering from the sample boundaries. This effect of boundaries is often treated via a simple specularity parameter. In the case of a sample in the nanoribbon geometry (width much shorter than length), the specularity parameter model for boundary scattering gives a simple and widely used approximation for the relaxation time \cite{Balandin_APL_09}: $
 \tau^{\text{b}}_{j,\vec{k}} = \frac{W}{v_{j,\vec{k}}} \frac{1+p}{1-p}$,
where $W$ is the ribbon width and $p$ is the specularity parameter.

In order to illustrate the importance of long-wave-length phonons for thermal transport in graphene samples, we show the contributions to room-temperature  thermal conductivity $\kappa$ [Eq. (\ref{equ:RTA})] by the phonons with different wave numbers $k$ [Fig. \ref{fig:NormalizedNumbers}(a1)] and different frequencies $f$ [Fig. \ref{fig:NormalizedNumbers}(a2)]; essentially, Figs.  \ref{fig:NormalizedNumbers}(a1) and (a2) are plots of the integrand from Eq. (\ref{equ:RTA}) if the integral is over $k$ or over $f$, respectively. We assume standard umklapp and isotope scattering rates that are quadratic in frequency \cite{Knezevic_APL_11} and isotropic phonon dispersions that fit the full dispersions quite well and contain both linear and quadratic terms \cite{Knezevic_JAP_14}. The scattering lifetime of long-wave-length phonons has an upper bound due to scattering from the sample boundaries, and we assume nanoribbon geometry with $W=25\,\mathrm{nm}$ and a constant specularity parameter of 0.9 \cite{Balandin_PRB_09}; this scattering process  will be important only at the center of the Brillouin zone, for very long wave lengths, otherwise it will be overshadowed by phonon--phonon and isotope scattering.

To quantify the range of important wave lengths for thermal transport in graphene, we seek the maxima of the curves in Figs. \ref{fig:NormalizedNumbers}(a1) and \ref{fig:NormalizedNumbers}(a2). The maxima occur at wave numbers 3.37 and 5.28 1/nm (wave lengths of 1.86 and 1.19 nm) for longitudinal acoustic (LA) and transverse acoustic (TA) phonons, respectively, and are marked with black-orange and black-blue vertical lines throughout Fig. \ref{fig:NormalizedNumbers}. These $k$'s correspond to frequencies of 9.33 and 11.0 THz, respectively. The peak wave numbers in Fig. \ref{fig:NormalizedNumbers}(a1) can be considered as roughly the midpoints of the ranges containing the most important wave vectors for thermal conduction; these ranges coincide with the ranges of nearly isotropic dispersions and the continuum limit [see Figs. \ref{fig:NormalizedNumbers} (b1) and (b2) for dispersions of TA and LA phonons, respectively, where the constant-wave-number circles corresponding to the peaks from Fig. \ref{fig:NormalizedNumbers}(a1) are denoted in black-blue and black-orange; see also density of states from full dispersion in Fig. \ref{fig:NormalizedNumbers}(c1) and the phonon number per unit frequency in Fig. \ref{fig:NormalizedNumbers}(c2)].

Overall, based on full dispersions and DOS in graphene, we can conclude that the range containing the wave vectors that are the highest contributors to thermal conductivity has considerable overlap with the range in which the approximation of an isotropic elastic solid holds (i.e., the continuum limit). 

As an aside, we note that, even though there are relatively few phonons in the continuum limit in 3D materials, elastic wave scattering remains useful for understanding thermal transport in these systems in the presence of disordered boundaries. Wave scattering from rough, random surfaces has universal features such as randomizing the direction of the outgoing energy. For that reason, early models of phonon--surface scattering were adapted from electromagnetic wave scattering \cite{Casimir_38, Ziman_PRS_55}, and many newer phonon--surface scattering models are based on scalar and elastic waves \cite{Cross_PRL_01, Jang_JAP_12, Lee_APL_12, Lambert_PRB_99, Cross_PRB_01, Lifshitz_PRB_01, Yung_PRB_01, Vasilopoulos_PRB_01, Geller_PRB_04, Xie_JoPD_07, JGong_PRB_09}.

\subsection{Our model system}\label{sec:our_system}

Because our system is classical, if we truly modeled a continuum material we would suffer from a problem analogous to the ``ultraviolet catastrophe'': there would be $k_BT$ energy in each of an infinite number of modes, leading to an infinite energy in our system. One way to avoid this problem is to set a cutoff wave length and only consider modes with longer wave lengths. This happens naturally when we discretize a continuum material using the FDTD method (Sec. \ref{sec:discretization}). To ensure that our model has physical relevance, we choose our cutoff wave length to be that in graphene. We accomplished this by setting the grid spacing in our system to be the graphene lattice constant (Sec. \ref{sec:FDTD}).

The benefits of our choice of grid spacing can be seen in the right half of Fig. \ref{fig:NormalizedNumbers}, which shows the density of states [Fig. \ref{fig:NormalizedNumbers}(d1)] and phonon number per unit frequency [Fig. \ref{fig:NormalizedNumbers}(d2)] in our system. Although our system has fewer continuum phonons than graphene due to Maxwell-Boltzmann statistics [contrast Fig. \ref{fig:NormalizedNumbers}(c1) and Fig. \ref{fig:NormalizedNumbers}(d2)], most of the phonons in our system still fall in the continuum limit. Also note the similarities between the cutoff frequencies and Van Hove singularities between graphene [Fig. \ref{fig:NormalizedNumbers}(c1)] and our graphenelike system with a grid cell size equal to the graphene lattice constant [Fig. \ref{fig:NormalizedNumbers}(c2)]. Without appropriate discretization our system would not have these similarities to graphene \cite{Mingo_PRB_10}.

We emphasize that our graphenelike model system is not intended for a direct and quantitative simulation of real graphene samples; e.g., the model system is harmonic, while graphene has anharmonicity. Still, our model system has many common features with graphene. For instance, Figs. \ref{fig:NormalizedNumbers}(b1)--(b4) show the full dispersion relations for graphene \cite{Knezevic_JAP_14} in panels (b1) and (b2) versus those obtained for our graphenelike FDTD simulation [Eq. (\ref{equ:dispersion})]. The graphene and FDTD dispersion relations are in good agreement up to the $k$'s of maximum contribution to thermal conductivity in the diffusive regime; the agreement worsens with higher $k$, which is particularly pronounced for TA phonons (although the DOS remains in good agreement for TA phonons). Finally, we see in Fig. \ref{fig:NormalizedNumbers}(d2) that our FDTD simulation has a sizeable population of phonons at and above the frequency of maximum contribution to thermal conductivity.

In short, while our FDTD simulation neither truly models atomistic graphene nor a perfect continuum 2D material, it is a good hybrid that provides useful insight into both.

\subsection{Boundary conditions}\label{sec:BCs}

At the surface, we consider both fixed (Dirichlet) and free (Neumann) boundary conditions (BCs). Let $\hat{n}$ be the surface normal vector. For elastic waves, the fixed BC is $\mathbf{u} = 0$ and the free BC is $\mathbf{\tau}\hat{n}=0$. For scalar waves, the fixed BC is $\phi = 0$ and the free BC is $\nabla \phi \cdot \hat{n} = 0$.

Free-standing nanostructures (such as suspended graphene nanoribbons) have unrestrained surfaces, which are equivalent to free BCs. We also analyze fixed BCs, because fixing the boundaries enables us to ``turn off'' Rayleigh waves, which exist only for elastic waves near a free surface. (Studying fixed BCs may also be useful for understanding nanostructures with edges that are not entirely free, such as in supported graphene nanoribbons.)

\subsection{Rayleigh waves}\label{sec:Rayleigh}

There are many types of surface waves in atomic and inhomogeneous elastic materials \cite{Wallis_PSS_74, Maradudin_NPD_87, Wallis_SS_94}. Graphene nanoribbons are no exception and support a number of in-plane and out-of-plane edge modes, with properties that depend on the terminating atoms \cite{Savin_PRB_10}. (Thermal transport in supported graphene will also be affected by Rayleigh waves in the substrate \cite{Volz_PRB_14}; the substrate Rayleigh waves result in out-of-plane motion of the nanoribbon and are not considered here.)

For the uniform elastic materials we consider here, Rayleigh waves (Fig.  \ref{fig:Rayleigh_wave}) are the only type of surface wave \cite{Maradudin_NPD_87}. Here, we briefly review some important facts about Rayleigh waves \cite{Rayleigh_PLMS_1885, Landau_Elasticity}.

The general form for a Rayleigh wave propagating in the $x$-direction near the surface of a semi-infinite bulk material ($y\geq 0$) is given by

\begin{align}\label{equ:Rayleigh_wave}
    \mathbf{u}\left(x, y, t\right) &= e^{i (k x-\omega t)}\left[\left(a \gamma_t e^{-\gamma_t y} + b k e^{-\gamma_l y}\right)\uvec{x} \right.\nonumber\\
    & \left. {} - i \left(a k e^{-\gamma_t y} + b \gamma_l e^{-\gamma_l y}\right)\uvec{y} \right],
\end{align}

\noindent
where $k$ is the wave number, $\omega$ is the angular frequency, $a$ and $b$ capture the amplitude of the wave at the surface, and $\gamma_l$ and $\gamma_t$ capture the exponential decay of amplitude away from the surface. Rayleigh waves result in a ``rolling'' motion, where $\mathbf{u}$ rotates 360 degrees each period.

Rayleigh waves have a linear dispersion relation, $\omega = c_r k$, where $c_r$ is the Rayleigh-wave group velocity. $c_r$ has a complicated dependence on the Lam\'e parameters. For $\lambda>0$, $0.874 c_t < c_r < 0.955 c_t$. For the graphenelike Lam\'e parameters we use here (see Sec. \ref{sec:FDTD}), $c_r=0.89c_t$, $\gamma_l = 2.19 k, \, \gamma_t = 3.85 k$, and $a=1.32 b$.

Rayleigh waves are slower than transverse or longitudinal waves, extend into the structure a distance comparable to their wave length, and can only exist with free BC because they have a nonzero displacement at the surfaces. Since a Rayleigh wave is a type of elastic wave, it can be decomposed into two scalar waves. The two scalar waves that represent a Rayleigh wave are not independent, but are coupled together by the free boundary condition \cite{Brekhovskikh_Book}. Applying the free boundary conditions separately to the two scalar waves ($\nabla \phi \cdot \hat{n} = 0$, $\nabla \psi \cdot \hat{n} = 0$) is \textit{not} equivalent to the free boundary condition for an elastic wave ($\mathbf{\tau}\hat{n}=0$). While two interdependent scalar waves are needed to form a Rayleigh wave, most existing scalar phonon--surface scattering models only use a single scalar wave \cite{Lambert_PRB_99, Cross_PRB_01, Lifshitz_PRB_01, Yung_PRB_01, Vasilopoulos_PRB_01, Geller_PRB_04, Xie_JoPD_07, JGong_PRB_09} and thus do not capture Rayleigh waves.

\section{Numerical Method}\label{sec:method}

\subsection{The finite-difference time-domain (FDTD) method for the elastic and scalar wave equation}\label{sec:FDTD}

We solve the elastic (\ref{equ:elastic_wave}) and scalar (\ref{eqn:scalar_waves}) wave equations in our structures using the finite-difference time-domain method, which is an efficient technique that discretizes the wave equation in both space and time by replacing the partial derivatives with finite differences. While the FDTD method is best known for solving the electromagnetic wave equation \cite{Yee_ITAP_66, Taflove_FDTD}, the method has been used with many wave equations, such as the Schr\"odinger \cite{Harmuth_JoMaP_57, Sullivan_JAP_12},  Klein-Gordon \cite{Harmuth_JoMaP_57}, scalar \cite{Virieux_GeoP_84}, and elastic wave equations \cite{Alterman_BSSA_68, Virieux_GeoP_86}. Elastic-medium FDTD has been used to investigate transmission through superlattices \cite{Tamura_JoPCM_97} and phononic materials \cite{Garcia_JAP_00, Garcia_PRL_00, Prevost_PRL_01}, but it had not been used before to calculate thermal conductivity.

Unlike many methods used to investigate elastic materials, elastic-medium FDTD is not limited to weak roughness or to any specific geometry. The technique is computationally simple and fast; the core of the simulation requires only a few lines of element-wise array operations, which can be computed quickly on modern processors.

Our FDTD method faithfully reproduces elastic waves with wave lengths greater than about 10 grid cells \cite{Levander_GeoP_88}, but gradually becomes less accurate at shorter wave lengths, which are already outside the continuum limit of atomic solids.

We use graphenelike Lam\'e parameters from \cite{Fasolino_PRL_09}: $\lambda = 32.0 \; \mathrm{J/m^2}$ and $\mu = 160.2 \; \mathrm{J/m^2}$. (Equivalently, $c_l = 2.14\times10^4 \; \mathrm{m/s}$ and $c_t = 1.44\times10^4 \; \mathrm{m/s}$.) We choose the grid-cell size (GCS), denoted $h$, to be the graphene lattice constant ($h$=0.246 nm), which means that the shortest wave length in our system will be similar to that in graphene. We use the GCS $h$ as a unit of length in this paper. Our choice of material parameters and $h$ sets the stability condition for the simulation time step $\Delta t < h/c_l \sqrt{2}$ \cite{Virieux_GeoP_86}. We chose $\Delta t = 0.95$ fs, one tenth of the maximum allowed value. Our time step is comparable to the 0.1-0.5 fs often used in graphene MD simulations \cite{Li_PRB_09, XGong_APL_09, YChen_NLet_09, Keblinski_APL_10, Dames_JAP_12}. We simulate 100-GCS-wide (24.6 nm) nanoribbons with random surface roughness that has a Gaussian autocorrelation function $C_g\left(x\right) = \Delta^2 e^{-x^2/\xi^2}$, where $\Delta$ and $\xi$ are the rms roughness and correlation length, respectively. The details for generating these random surfaces are given in Ref. \cite{Knezevic_PRB_12}.

While the elastic wave equation (\ref{equ:elastic_wave}) can be directly solved with the FDTD method to obtain $\mathbf{u}$ \cite{Alterman_BSSA_68}, we instead use the velocity-stress formulation \cite{Virieux_GeoP_86} because it allows for a simple and stable implementation of free BCs and because the velocity and stress are ultimately what we need to find $\mathbf{J}$ [Eq. (\ref{eq:J})].

Taking the time derivatives of the strain equation (\ref{equ:strain}), Hooke's law (\ref{equ:Hooke}), and elastic wave equation (\ref{equ:elastic_wave}) while defining $\mathbf{v} = \dot{\mathbf{u}}$ yields:

\begin{subequations}\label{equ:velocity-stress}
  \begin{align}
        \dot{\sigma}_{ij} &= \frac{1}{2}\left(\partial_j v_i + \partial_i v_j\right), \\
        \dot{\tau}_{ij} &= \lambda \dot{\sigma}_{ll} \delta_{ij} + 2\mu\dot{\sigma}_{ij}, \\
        \rho{\dot{\mathbf{v}}} &= \mathbf{\nabla} \cdot \mathbf{\tau}.
    \end{align}
\end{subequations}

\begin{figure}
 \centering
 \includegraphics[width=3 in]{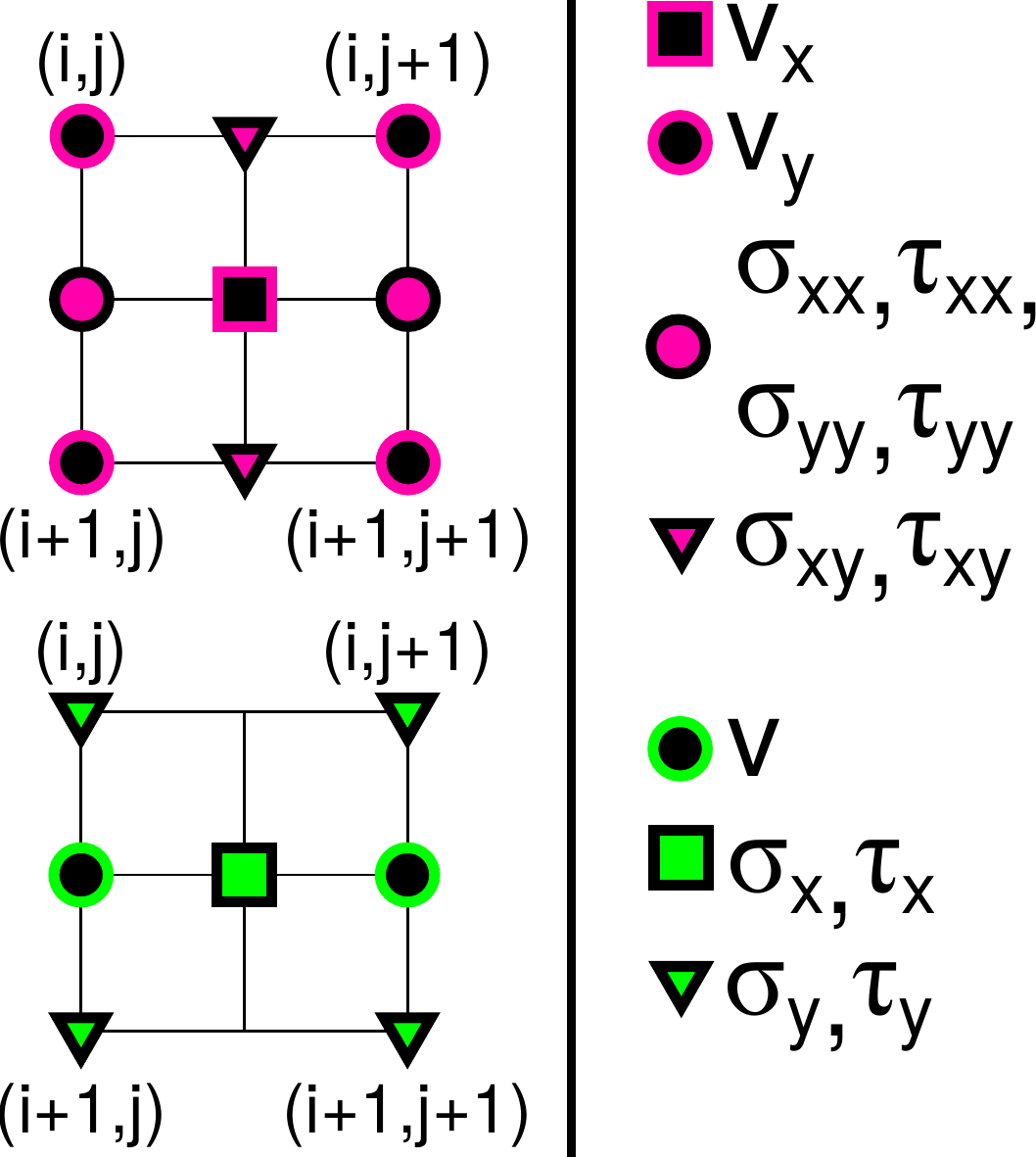}
 \caption{The staggered grid used for the FDTD solution to the  elastic (top) and scalar (bottom) wave equations. $(i,j)$ enumerate the grid cells along the $x$ and $y$ directions. The symbols (squares, triangles, and circles) show where on the grid the different components of $\mathbf{v}$, $\mathbf{\sigma}$, and $\mathbf{\tau}$ are defined. } \label{fig:grid}
\end{figure}

\noindent The elastic wave equation is thus broken into two first-order differential equations. [Although there are three equations above, $\dot{\mathbf{\sigma}}$ and $\dot{\mathbf{\tau}}$ are linearly related via Hooke's law (\ref{equ:Hooke}).] The velocity-stress formulation \cite{Virieux_GeoP_86} solves for the stress--strain and velocity using a leapfrog technique on a staggered, square grid (See Fig. \ref{fig:grid}). We found that the second-order-accurate \cite{Virieux_GeoP_86} and fourth-order-accurate \cite{Zeng_GeoP_12} spatial finite difference operators were both stable and suitable for our purposes. We used the second-order-accurate operator because it is computationally simpler.

To implement free boundaries, we used the ``vacuum formalism,'' where materials parameters $\mu$, $\lambda$, and $\rho^{-1}$ are set to zero outside the structure. For accuracy and stability, a half-grid-cell-thick fictitious layer of material is added around the structure to ease the transition from the material to vacuum \cite{Zeng_GeoP_12}. Fixed BCs are simpler to implement: we force $\mathbf{v} = 0$ on the surface.

The 2D scalar wave equation can be recovered by taking the 3D elastic wave equation and setting $\partial_z v_z = 0$. To keep our elastic and scalar FDTD methods consistent, we take the 3D generalization of our elastic wave FDTD technique \cite{Graves_BSSA_96} and set $\partial_z \dot{v}_z = 0$. The result of this mathematical convenience is a 2D scalar wave equation FDTD method. (The method can also be derived directly \cite{Virieux_GeoP_84}.) Free and fixed BCs are enforced by setting the stress or velocity, respectively, to zero on the surface.

Because we simulate a linear elastic material, our model does not include phonon-phonon scattering, which would imply a nonlinear restoring force in the wave equation \cite{Ziman_Book}. Our simulation is suitable for investigating structures where surface or boundary roughness is the dominant scattering mechanism.

\subsubsection{Energy conservation in long simulations}\label{sec:energy conservation}

The elastic wave FDTD technique was originally developed to model earthquakes \cite{Alterman_BSSA_68, Kelly_GeoP_76}. Many commonly used free-surface implementations are known to have long-term instabilities \cite{Stacey_BSSA_94}; however, these instabilities are not an issue for typical seismic simulations, which are often short and have absorbing boundary conditions along some domain edges. In contrast, our simulations require  stability and energy conservation even for very long simulations (millions of time steps). It had not been previously reported that the combination of BCs and FDTD method we report here is energy-stable \cite{Tsoflias_email}.

\subsubsection{Effects of discretization}\label{sec:discretization}

Discretizing wave equations introduces dispersion and anisotropy to the dispersion relations. Dispersion and anisotropy are undesirable for modeling continuum materials. However, as we ultimately care about phonons in atomic materials, which also have dispersion and anisotropy, discretization can actually bring  the elastic medium model closer to an atomic model. The dispersion relation for our FDTD method is \cite{Virieux_GeoP_86}:

\begin{equation}\label{equ:dispersion}
 \omega_i = \frac{2}{\Delta t} \arcsin{\left( \frac{c_i \Delta t}{h}\sqrt{\sin^2\frac{h k_x}{2} + \sin^2\frac{h k_y}{2}} \right)},
\end{equation}

\noindent
where $\vec{k}$ is the wave vector, and $i$ can be either $l$ or $t$ for longitudinal or transverse phonons, respectively. Dispersion relations of this form are common to other FDTD methods \cite{Taflove_FDTD}.

The argument of the $\arcsin$ is small because $\frac{c_i \Delta t}{h}$ in our system equals 0.07 and 0.05 for longitudinal and transverse phonons, respectively, so we can approximate $\arcsin{(x)} \approx x$ and get a dispersion relation very similar to the classic dispersion relation for a periodic one-dimensional lattice with harmonic potentials \cite{Ziman_Book, Ashcroft_Mermin}: $\omega \propto \left|\sin \left(h k/2\right)\right|$. In the limit of small $|\vec{k}|$, Eq. (\ref{equ:dispersion}) becomes the dispersion relation for continuum materials: $\omega_i = c_i |\vec{k}|$.

We also note that it is possible to explicitly add anisotropy to elastic-material FDTD simulations at the cost of additional complexity \cite{Riollet_GeoP_95, Sato_JJAP_08}. It should also be possible to implicitly change the anisotropy by choosing grids with different geometries. For example, a hexagonal grid \cite{Taflove_FDTD} should lead to a dispersion relation with hexagonal symmetry.

Discretization also leads to minimum wave lengths for waves in the system. For bulk waves (transverse and longitudinal), the shortest wave length is two grid cells ($2h$) with displacements that alternate $\rightarrow, \leftarrow$ or $\uparrow, \downarrow$ for longitudinal and transverse waves, respectively, that propagate in the horizontal direction. However, Rayleigh waves correspond to a ``rolling'' motion (Sec. \ref{sec:Rayleigh}) which cannot be captured with just two grid cells. Thus, the shortest wave length for Rayleigh waves we capture is four grid cells ($4h$), with displacements that cycle $\uparrow, \leftarrow, \downarrow, \rightarrow$ for a wave propagating in the horizontal direction. In the absence of dispersion, this minimal wave length corresponds to a cutoff frequency of $\sfrac{c_r}{4h}\approx 13$ THz for Rayleigh waves. Other frequencies that will be important later correspond to the shortest wave length for waves propagating at a 45 degree angle from the ribbon axis. That corresponds to a wave length of $2^{\sfrac{3}{2}}h$ and frequencies of $\frac{c_l}{2^{\sfrac{3}{2}}h}\approx 31$ THz and $\frac{c_t}{2^{\sfrac{3}{2}}h}\approx 21$ THz for longitudinal and transverse waves, respectively.

\subsubsection{FDTD vs. Molecular Dynamics}\label{sec:MD_comparison}

Because of the discretization, our FDTD model has similarities to a molecular dynamics simulation, with springlike nearest-neighbor potentials and with the discretized material elements playing a similar role to atoms in MD. The primary advantages of an FDTD simulation over an MD simulation are simplicity, scalability, and computational speed. We can simulate relatively large structures with significantly less computational cost than an MD simulation. The trade-off is that the FDTD method cannot accurately account for the short-wave-length limit or anharmonic potentials of atomic materials. The harmonic potentials and classical statistics also mean that our thermal conductivity is independent of temperature.

\subsection{Thermal conductivity calculation}\label{sec:kappa}

\begin{figure}
 \centering
 \includegraphics[width=3 in]{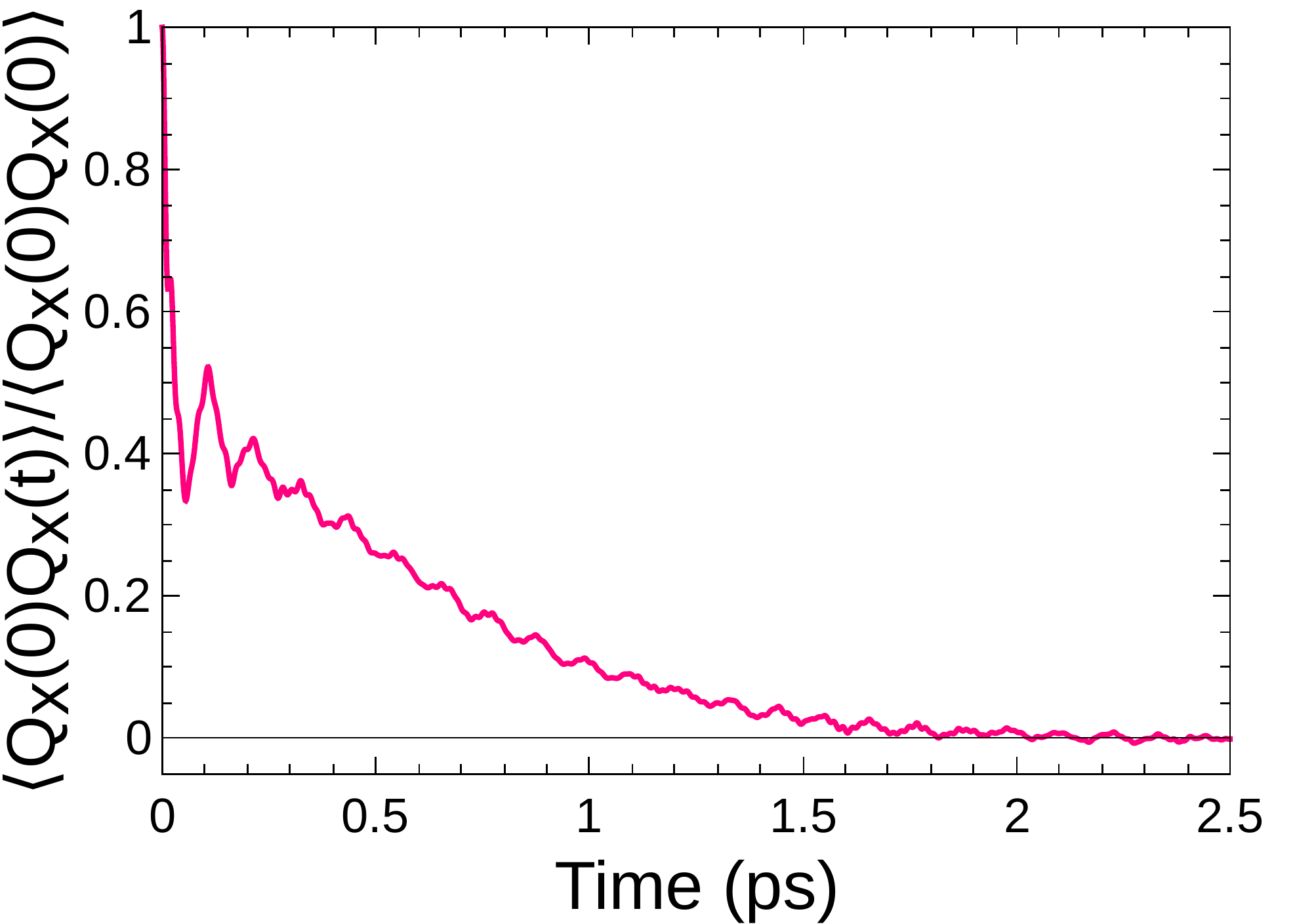}
 \caption{Representative heat current autocorrelation function versus time. The curve is similar to those obtained for equilibrium molecular dynamics simulations \cite{Yip_JNucM_98}.} \label{fig:hcacf}
\end{figure}

We compute the thermal conductivity in a similar way to an equilibrium molecular dynamics simulation \cite{Hoover_PRB_86, Yip_JNucM_98}. First, we initialize all the discretized elements with a random velocity drawn from a Maxwell-Boltzmann distribution. The simulation can become unstable if the net momentum is too far from zero. Because we draw a finite number of samples from the Maxwell-Boltzmann distribution, the net momentum will be small but nonzero. To account for this, before starting the simulation we adjust the velocities so that the structure has zero net momentum. We likewise adjust the velocities so that the kinetic energy is equally divided between the $x$ and $y$ motion, $x$ being along the nanoribbon. All structures in this paper have periodic boundary conditions in the $x$ direction (along the axis of the ribbon), so energy neither leaves nor enters the simulation.

We run the simulation for 100,000 time steps to let the system equilibrate. Then, the simulation runs an additional 900,000 time steps in equilibrium. At each time step, we calculate the heat current $\mathbf{Q}$, which is the spatial integral of the heat current density $\mathbf{J}$ [Eq. (\ref{eq:J})]. Finally, we calculate the thermal conductivity $\kappa$ using the Green-Kubo formula:

\begin{equation}
    \kappa = \frac{1}{k_B T^2 \Omega}\int_{0}^{\infty} \left< Q_x\left(0\right) Q_x\left(t\right) \right>dt \, ,
\end{equation}

\noindent
where $T$ is the system temperature, $\Omega$ is the system volume, and the $x$-axis is along the ribbon. To calculate the system volume, we use the standard technique of multiplying the surface area by graphite interplanar distance of 0.335 nm \cite{Keblinski_APL_10}. We directly compute the integral and cut it off after the first dip, when the heat current autocorrelation function $\left< Q_x\left(0\right) Q_x\left(t\right) \right>$ first reaches zero \cite{Yip_JNucM_98}. Figure \ref{fig:hcacf} shows that the heat current autocorrelation function obtained via FDTD has a temporal dependence similar to that obtained in MD simulations, underscoring the similarity between the two techniques.

\section{Surface scattering and mode conversion}\label{sec:mode_conversion}

\begin{figure*}
\centering\includegraphics[width=6 in]{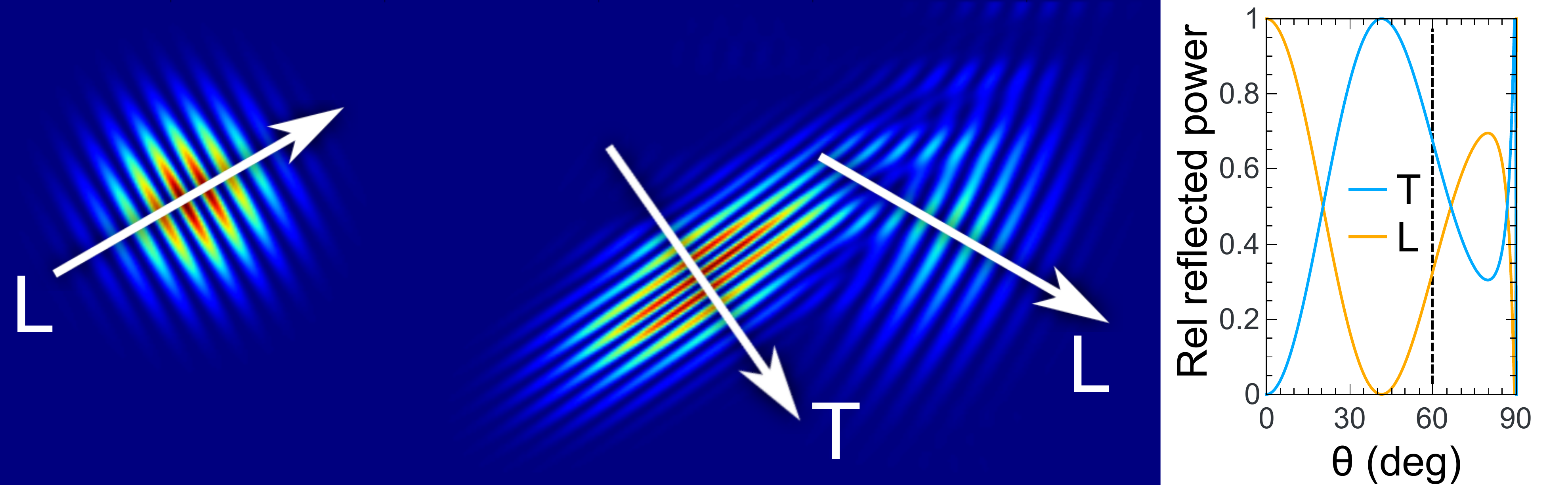}
\caption{(Left panel) Visualization of elastic wave mode conversion at a smooth surface.  Color represents the spatial profile of the energy density (arbitrary units; red--high, blue--low). A longitudinal wave packet is incident on a free, smooth top surface at $60^\circ$ from the surface normal. One longitudinal and one transverse wave packet are reflected. The transverse wave can be identified by its shorter wave length and slower group velocity. The plot to the right shows the relative energy in the scattered longitudinal (L) and transverse (T) wave packets as a function of the angle of incidence. The material parameters have a strong effect on the angular dependence seen in the plot.}
 \label{fig:smooth_mode_conversion}
\end{figure*}

The longitudinal and transverse phonon modes are independent of each other in an infinite medium. However, the two types of modes, and their corresponding scalar wave equations [Eq. (\ref{eqn:scalar_waves})], are not independent at a surface. For example, the fixed BC for the elastic wave equation ($\mathbf{u} = 0$) combined with Eq. (\ref{equ:decompostition}) requires $\nabla\phi = -\nabla\times\mathbf{\Psi}$. So, the fixed BC for the elastic wave equation is not equivalent to independently applying fixed boundary conditions ($\phi=0$ and $\psi=0$) to the two scalar wave equations. In short, the elastic wave equation allows for mode conversion at the surface, while the scalar wave equations do not allow for mode conversion if their boundary conditions are applied independently. Existing scalar wave models of phonon--surface scattering typically employ only a single scalar wave \cite{Lambert_PRB_99, Cross_PRB_01, Lifshitz_PRB_01, Yung_PRB_01, Vasilopoulos_PRB_01, Geller_PRB_04, Xie_JoPD_07, JGong_PRB_09}, which precludes mode conversion.

Figure \ref{fig:smooth_mode_conversion} illustrates mode conversion at the surface via a colorized energy-density profile for a simple example: a longitudinal wave packet incident upon a smooth, free surface at the top. The longitudinal wave is reflected into longitudinal and transverse waves; there is no conversion from bulk into Rayleigh waves at a smooth surface \cite{Landau_Elasticity, Wolfe_Book, Klemens_PRL_70, Sakuma_PRB_73, Maznev_PRB_15}. Wei \textit{et al.} \cite{Dames_JAP_12} obtained similar results based on a molecular-dynamics simulation of mode conversion between bulk modes at smooth graphene surfaces. There are many fine points of conversion between bulk modes, such as the angles where the incoming longitudinal wave is converted entirely into a reflected transverse wave (right panel in Fig. \ref{fig:smooth_mode_conversion}). However, in Sec. \ref{sec:results} we will show that mode conversion between bulk modes has little impact on phonon thermal transport in our nanoribbonlike systems and we therefore focus on conversion between bulk modes and Rayleigh waves at rough surfaces. (We note that conversion between bulk modes and Rayleigh waves has also been simulated with finite-element methods \cite{Schmidt_JAP_09, Schmidt_Ultra_13_I, Schmidt_Ultra_13_II}.)

\begin{figure}
 \centering
 \includegraphics[width=3 in]{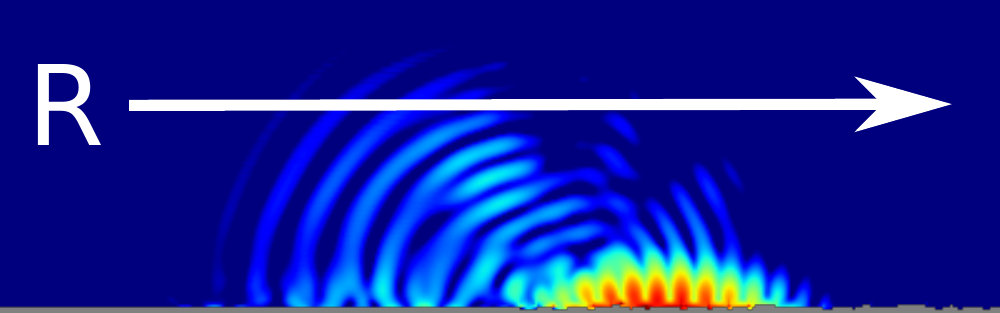}
 \caption{Snapshot of a Rayleigh wave that scattered from a rough surface. Color represents the energy-density profile (log scale, arbitrary units; red--high, blue--low). A Rayleigh wave packet was launched from left to right, moving first along a smooth bottom surface (left) and then along a rough bottom surface (right). Once the packet reaches the roughness (snapshot was taken shortly thereafter), it starts radiating energy into bulk modes. However, the conversion from Rayleigh into bulk modes is relatively weak. The energy leaves the packet slowly, and the packet continues to the right with little distortion.}
 \label{fig:Rayleigh_scattering}
\end{figure}

Figure \ref{fig:Rayleigh_scattering} shows a snapshot of the spatial energy density, represented by color, of a Rayleigh wave packet that moves from left to right along the bottom surface. The packet arrived from a region near a smooth surface (bottom left) and impinged upon a rougher region (bottom right); the snapshot was taken shortly thereafter. The packet remained largely intact, albeit slightly distorted, as it continued to travel along the rough surface. Some energy is radiated into bulk modes, but the amount is small compared to the amount of energy still in the packet (note that the energy density is plotted on a log scale). This finding is in line with previous studies of Rayleigh-wave scattering from disordered surfaces, which showed that Rayleigh waves were tolerant of material density disorder on scales smaller than the wave length \cite{Klemens_PRL_70, Sakuma_PRB_73}.

\begin{figure*}
 \centering
 \includegraphics[width=6 in]{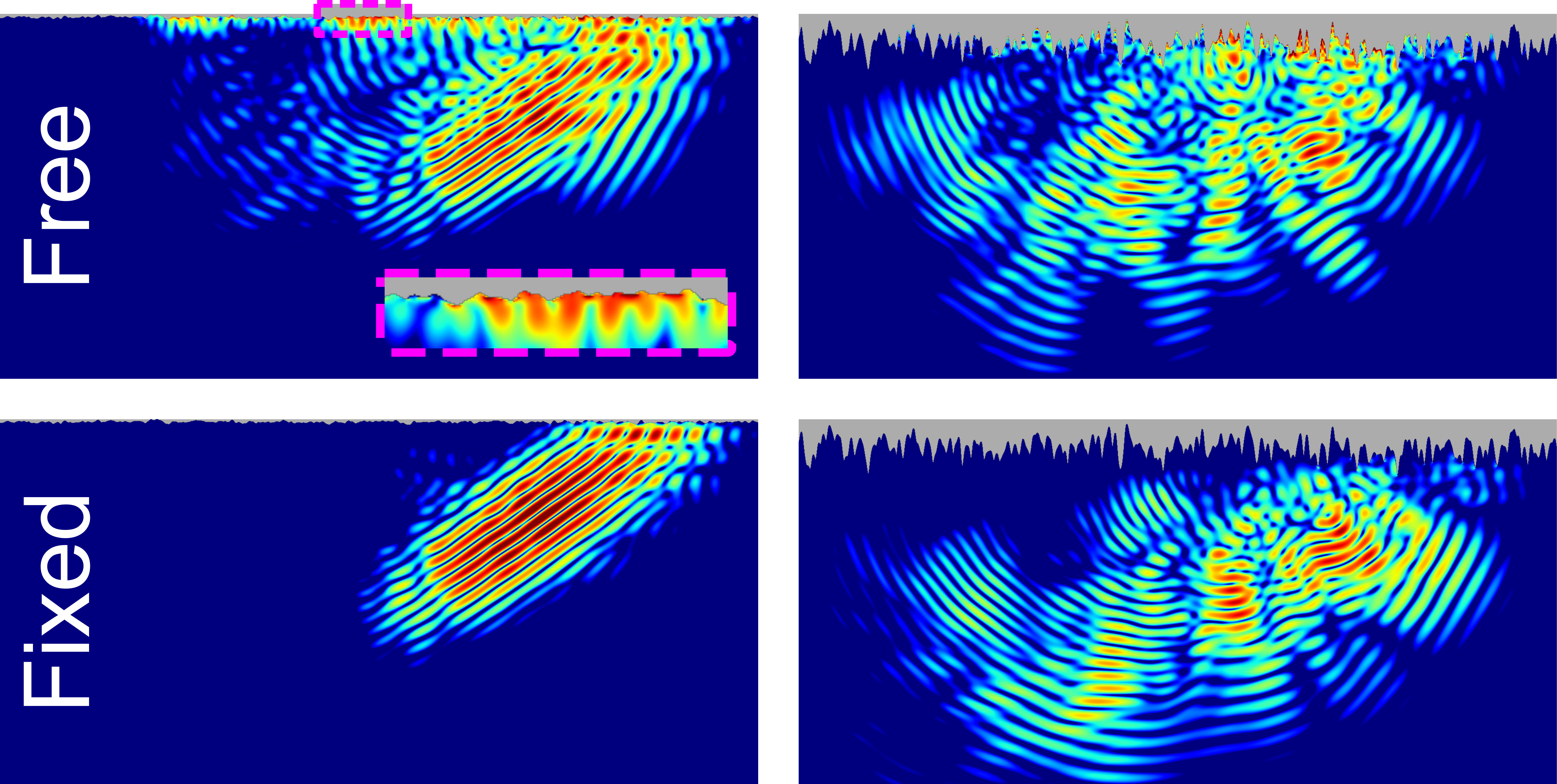}
 \caption{Snapshots of bulk elastic waves scattered from rough surfaces (top of each panel). Color represents the energy-density profile (log scale, arbitrary units; red--high, blue--low). A longitudinal wave packet was  incident on a surface that is nearly smooth and free (top left panel), very rough and free (top right), nearly smooth and fixed (bottom left), and very rough and fixed (bottom right). Surface Rayleigh modes are visible only for the free surfaces (i.e., energy is localized near the free surfaces), because fixed surfaces do not support surface modes. (Inset to top left panel) Zoom-in on the energy-density  profile for a free, nearly smooth surface (region inside the dashed box on the main panel) reveals a wave packet localized near the surface: a Rayleigh mode. (Compare with the Rayleigh wave packet in Figure \ref{fig:Rayleigh_scattering}. The relative size of the rms roughness $\Delta$ and packet wave length $\lambda$ are $\Delta/\lambda=0.025$ for the left two figures, $\Delta/\lambda=0.25$ for the right two figures. $\xi/\lambda=0.5$ for all figures ($\xi$ is the correlation length).}
 \label{fig:scattering}
\end{figure*}

Figure \ref{fig:scattering} shows examples of mode conversion for an incident longitudinal wave packet scattering from surfaces with different roughness (nearly smooth: left column, very rough: right column) and boundary conditions (free BC: top row; fixed BC: bottom row). Each panel shows the spatial energy density, represented by color and on the log scale. The incoming wave profile is the same as in Fig. \ref{fig:smooth_mode_conversion}, and is omitted here for clarity; the color represents only the energy density for the outgoing wave packets.  First, the specularly scattered bulk modes are visible, albeit distorted, for the nearly smooth surfaces of both BCs. In contrast, scattering from the very rough surfaces of both BC is very diffuse. Second, no energy remains localized at the fixed surfaces, because fixed surfaces do not support surface modes. In contrast, a significant amount of energy is captured near the free surfaces. The longitudinal wave incident on the nearly smooth free surface is partially converted into Rayleigh waves. One Rayleigh-wave energy-density profile has been enlarged in the inset to the top left panel.

For the very rough free surface (Fig. \ref{fig:scattering}, top right panel), the energy remaining near the surface is concentrated in a few places and has not propagated nearly as far along the surface as it did in case of its nearly smooth counterpart. The concentrated energy is similar to the spatially localized modes (SLMs) seen in molecular dynamics simulations \cite{Savin_EPL_09, Estreicher_PRB_14, Estreicher_JAP_15}. In particular, the surface modes we see are similar to the ``wag modes'' predicted for atoms terminating bonds on nanowires \cite{Estreicher_PRB_14, Estreicher_JAP_15} and the motion of molecular chains (``side phonon leads'') attached to the edges of nanoribbons \cite{Savin_EPL_09}. Instead of a terminating atom, a protuberance of the nanoribbon wags back and forth. (See Fig. \ref{fig:wag} and Sec. \ref{sec:VEA} for more detail.) SLMs were found to release their energy over relatively long time scales, and neighboring SLMs are often weakly coupled, so energy travels very slowly along the surface. Although the energy at the very rough free surface has effectively become localized and the wave is no longer a propagating one, for consistency, we will still refer to the localized surface modes as Rayleigh waves.

\section{Localization and vibrational eigenmode analysis}\label{sec:VEA}

To directly investigate localization, we preform a vibrational eigenmode analysis (VEA) \cite{Feldman_PRB_93, Li_NLet_10, Russ_EPL_04} of our system. VEA involves writing the equations of motion for the system [Eq. (\ref{equ:velocity-stress})] in matrix form and finding all the modes of the system. VEA allows us to find the frequencies, spatial energy distribution, and motion of each mode. Because finding the eigenvectors and eigenvalues of a large matrix is computationally expensive, we only show results for ribbons 70 grid cells or less in width.

The localization of a mode is commonly quantified by its \textit{participation ratio}, $p$, \cite{Feldman_PRB_93, Li_NLet_10}:

$$p=\left(N \sum_i E_i^2\right)^{-1},$$

\noindent
where $N$ is the number of grid cells inside the ribbon, and $E_i$ is the is the time averaged kinetic energy of the $i$th grid cell. $E_i$ is found using VEA and is normalized so that $\sum_i E_i =1$. The participation ratio is usually defined in terms of the displacement of the grid cells \cite{Feldman_PRB_93} rather than the velocity (and thus average kinetic energy). However, for a harmonic oscillator, the amplitudes of the displacement and velocity are proportional to each other, and any differences in constants are removed by the normalization. Our definition of the participation ratio is equivalent to the standard definition. We use our definition because our FDTD method yields the velocity but not the displacement.

$p$ measures how evenly the energy is spatially distributed in the system. If the energy is evenly distributed among the grid cells ($E_i=N^{-1}$), then $p=1$ (total delocalization, such as a plane wave). If all the energy is in one grid cell, then $p=N^{-1}$ (total localization). Values between the two limits indicate varying degrees of localization.

Note that $p=1$ is impossible even inside a smooth ribbon because there must be standing waves in at least one direction. Standing waves have nodes where the displacement (and thus kinetic energy) is zero, so the kinetic energy cannot be evenly distributed. Consider scalar waves in a smooth ribbon of length $L$ and width $W$, and take $n,m$ to be natural numbers. A wave that is traveling along  $x$ and standing in $y$ will be of the form $z\left(x,y,t\right) \sim \sin\left(\pi n y /W \right) e^{i \left( k x-\omega t\right)}$, which results in $p = \sfrac{2}{3}$. A wave that is standing in both $x$ and $y$ directions will be of the form $z\left(x,y,t\right) \sim\sin\left(\pi n x /L \right) \sin\left(\pi n y /W \right) e^{i \omega t}$, which results in $p =\left(\sfrac{2}{3} \right)^2=\sfrac{4}{9}$ \cite{Russ_EPL_04}. Our structure has finite length but periodic boundary conditions, which means the modes are propagating but have longitudinal constraints stemming from periodicity as well as a finite grid cell size. Therefore, it is not surprising that the numerically obtained participation ratios for a smooth ribbon fall between $\sfrac{4}{9}$ and $\sfrac{2}{3}$.

\begin{figure}
 \centering
 \includegraphics[width=3 in]{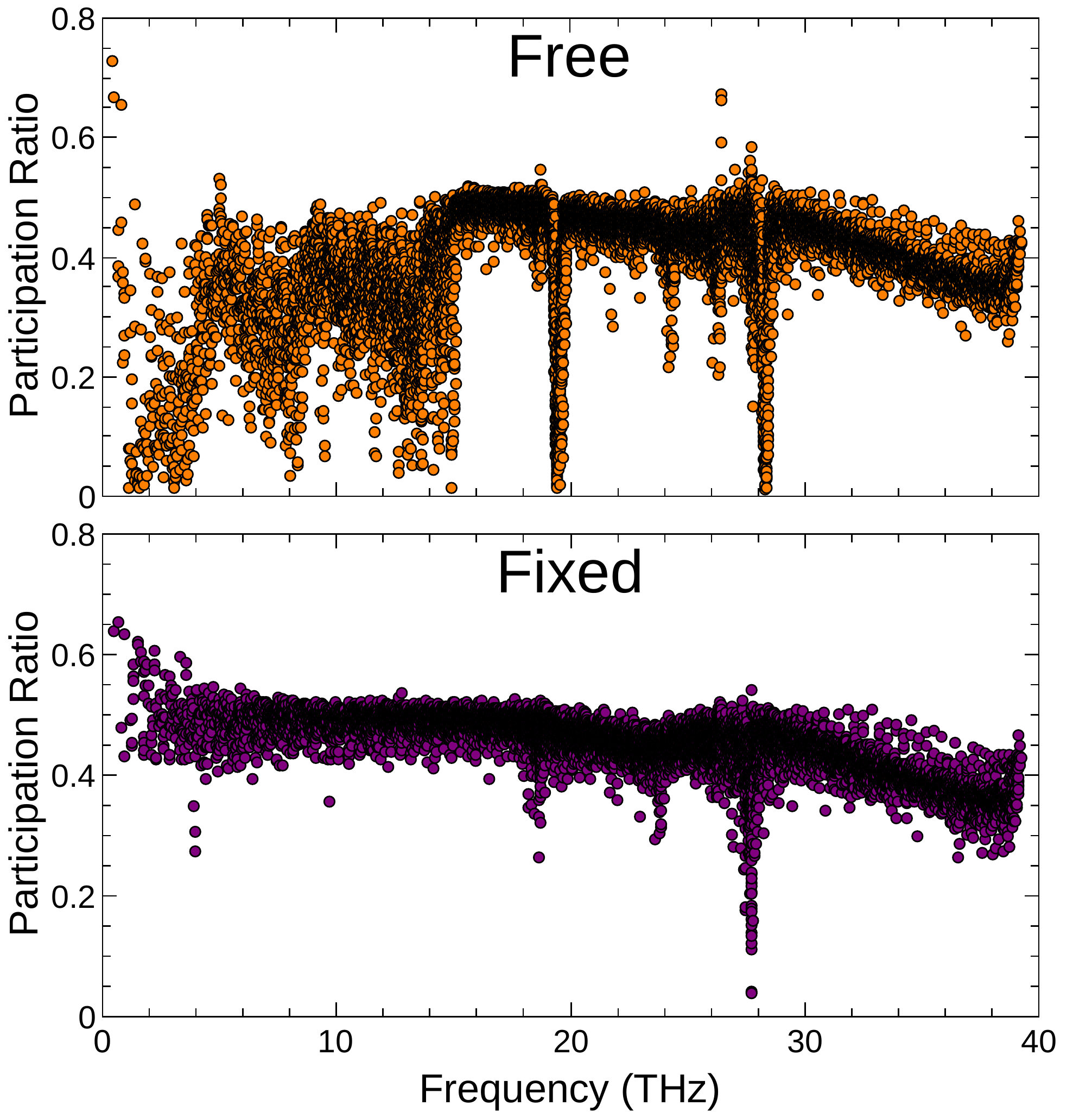}
 \caption{The participation ratio versus frequency for all modes of a $70h$-wide graphenelike nanoribbon with $\Delta = 2h, \xi=5h$, and free (top) and fixed (bottom) BCs. Free BC results in significant localization at low frequencies, where we expect to find Rayleigh waves. Traveling waves in a smooth ribbon will have a participation ratio of $\sfrac{2}{3}$. Indeed, all the calculated participation ratios are below $\sfrac{2}{3}$ except for one spurious point. $\sfrac{4}{9}$ is the smallest participation ratio possible in a smooth ribbon. Almost all the low frequency modes in the fixed ribbon have participation ratios above $\sfrac{4}{9}$, while almost all the low-frequency modes in the free ribbon have participation ratios below $\sfrac{4}{9}$. This clearly shows how free surfaces can increase the localization of all low-frequency modes; nearly all surface Rayleigh modes are in this range.} \label{fig:participation}
\end{figure}

The impact of boundary conditions on localization can be seen in Fig. \ref{fig:participation}. Free BCs results in significant localization at low frequencies, where we expect to find Rayleigh waves. As we noted in Sec. \ref{sec:continuum_limit}, low-frequency phonons are very common and important in actual 2D materials, so the localization of these phonons is particularly important. At higher frequencies, the participation ratios differ little between free and fixed BCs. While it is hard to definitively prove a connection between localization and conductivity, it is widely believed that localized modes contribute less to the conductivity than delocalized modes \cite{Galli_NLet_10, Li_NLet_10, Poulikakos_NLet_11, Henry_NJoP_16}.

We would like to stress that Fig. \ref{fig:participation} shows that almost all low-frequency modes (including bulk modes) localize when free BCs are introduced. Consider the frequency range below 15 THz. This is where we expect all Rayleigh waves to lie for structures with free BCs (Sec. \ref{sec:discretization}), although there are some small wag modes (described below) at higher frequencies. For fixed BCs (bottom panel) there are many modes below 15 THz, \textit{and they are all bulk modes} since there are no Rayleigh waves with fixed BCs. These low-frequency bulk modes should not disappear when we introduce free boundary conditions, so it must be that there are many bulk modes below 15 THz for ribbons with free BCs. However, it is clear that, when free BCs are introduced, almost all the modes below 15 THz have become more localized -- both the newly introduced Rayleigh waves and the preexisting bulk modes! The effect is striking. With fixed BCs, nearly all of these modes have participation ratios \textit{above} $\sfrac{4}{9}$. With free BCs, the vast majority of the modes now have participation ratios \textit{below} $\sfrac{4}{9}$. Recall participation ratios below $\sfrac{4}{9}$ are impossible for bulk modes in a smooth ribbon, so all participation ratios below the value $\sfrac{4}{9}$ are clear signs of localization due to disorder.

The large drop in participation ratio for free surfaces is due to surface roughness and is not an intrinsic effect of free BCs. Table \ref{table:particip_ratio} shows the average participation ratios for modes with frequencies below 15 THz as roughness is increased. For smooth ribbons, the average participation ratio is nearly the same for free and fixed BCs, so free BCs do not intrinsically lower the participation ratio. However, as the roughness is increased, the differences in the participation ratios become clear. The participation ratio for fixed ribbons drops slowly as roughness is increased, but the participation ratio for free ribbons drops precipitously. It is possible that this effect is related to mode conversion between Rayleigh and bulk waves (Sec. \ref{sec:mode_conversion}). Rayleigh-to-bulk mode conversion also requires free surfaces and is also nonexistent at smooth surfaces but becomes more prominent with increased roughness. Rayleigh-to-bulk mode conversion might lead to hybridized bulk-Rayleigh waves and lead to increased localization of the formerly bulk modes.

\begin{table}
\caption{Average participation ratio, $\bar{p}$, for modes with frequencies below 15 THz (which includes all Rayleigh waves) in graphenelike nanoribbons of width and length 70h and correlation length $\xi=5h$ at different rms roughness $\Delta$. Grid-cell size is $h=0.246$ nm. }
    \label{table:particip_ratio}
  \begin{tabular}{  l c c c c c }
    \hline\hline
    $\Delta \rightarrow$                     &  $0$ & $3h$  & $6h$ & $9h$ & $12h$\\ \hline
    $\bar{p}_{\mathrm{free}}$                         & 0.515 & 0.352 & 0.323 & 0.286 & 0.267 \\
    $\bar{p}_{\mathrm{fixed}}$                        & 0.543 & 0.493 & 0.483 & 0.474 & 0.465 \\
    $\frac{\bar{p}_{\mathrm{free}}}{\bar{p}_{\mathrm{fixed}}}$ & 0.948 & 0.714 &  0.668 & 0.602 & 0.573 \\
    \hline\hline
         \end{tabular}
\end{table}

\begin{figure*}
\centering\includegraphics[width=6 in]{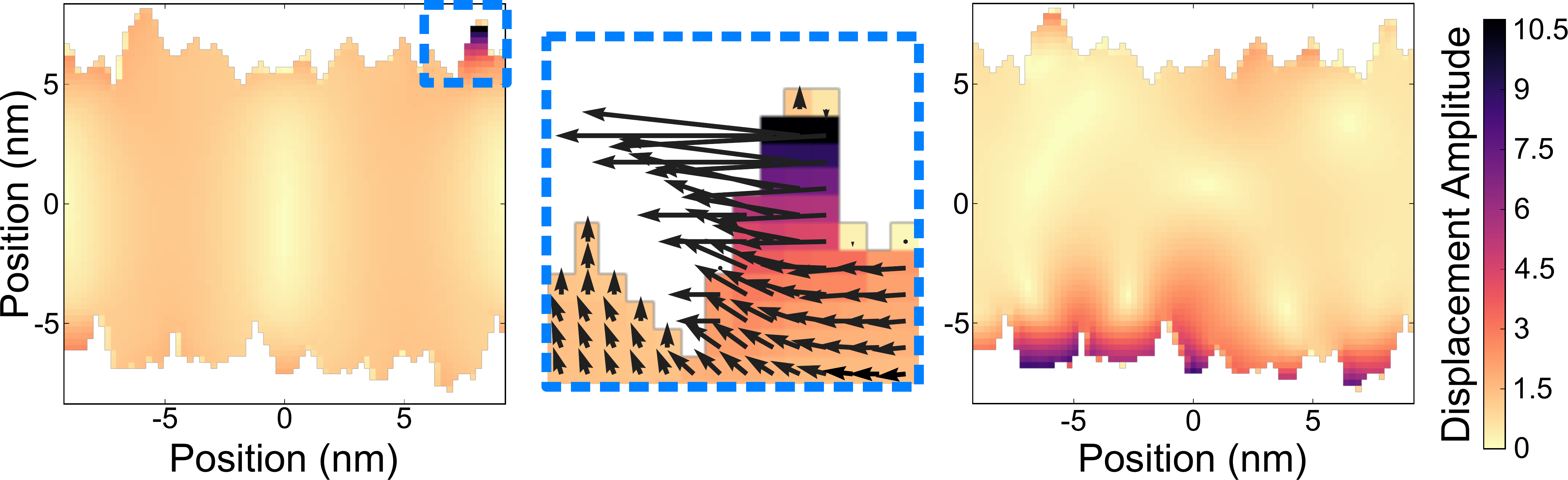}
\caption{Illustration of surface modes calculated with VEA for a $50h$-wide ribbon with $\Delta=3h$ and $\xi=5h$. Color is the relative amplitude of the velocity field (proportional to the square  root of kinetic energy) throughout the ribbon (colorbar to the right). (Left) A wag mode (participation ratio $p=0.089$): the energy is concentrated in a protrubance that ``wags'' side to side, as illustrated by (center) a close-up of the wag mode with a superimposed snapshot of the velocity field. (Right) A mode that is less localized than the wag mode but still concentrates energy at the surface.}
 \label{fig:wag}
\end{figure*}

VEA also allows us to visualize individual modes. Fig. \ref{fig:wag} shows one of the wag modes described in Sec. \ref{sec:mode_conversion} as well as a surface mode that is less localized. The less-localized surface mode looks like a series of wag modes interspersed with a couple gaps. These gaps suggest that the mode cannot propagate energy all the way through the structure. This could indicate that the structure is sufficiently rough to prevent Rayleigh waves from propagating. We have observed that the most localized modes are generally surface modes. That observation is consistent with the energy density distribution in the ribbon, which is discussed in Sec. \ref{sec:results} and Fig. \ref{fig:localization}. Fixed boundary conditions do not have a direct analog to wag modes because, unlike free boundary conditions, fixed boundary generally allow bulk modes to extend into surface irregularities so that the protuberances do not have their own modes \cite{Russ_EPL_04}.

The dips in Fig. \ref{fig:participation} that occur at approximately 20 and 30  THz represent ``bouncing ball'' modes that traverse the ribbon at 45 degree angles to the ribbon axis. Bouncing-ball modes are very short-wave-length modes that correspond the trajectory of point particles ``bouncing'' though the structure \cite{Stockman_book}. The frequencies of the dips correspond to the shortest possible wave length for longitudinal and transverse waves traveling at 45 degree angles to the ribbon axis (Sec. \ref{sec:discretization}). It is unclear why these bouncing ball modes are more localized than bouncing ball modes at other angles.

This means that the location of the dips (and some other interesting features in Fig. \ref{fig:participation}) are somewhat dependent on our choice of GCS. However, these features could also occur in atomic materials, where the lattice spacing is set by nature, not by theorists.

\section{Thermal conductivity and Rayleigh waves}
\subsection{Results}\label{sec:results}

We now turn to the thermal conductivity of nanoribbons. We model ribons that are 300 grid cells (73.8 nm) long and have an average width of 100 grid cells (24.6 nm). The length is sufficient to supress length effects (Sec. \ref{sec:length}). The results, summarized in Table \ref{table:results}, are the averages from 100,000 simulations. For each ribbon, we simulate the elastic wave equation and the two decoupled scalar equations. The thermal conductivities given for the scalar waves are the sum of the thermal conductivities of the two decoupled scalar waves. The results follow the general pattern seen elsewhere \cite{Yang_NLet_12, Knezevic_APL_15} of $\kappa$ decreasing with increasing rms roughness $\Delta$ and decreasing correlation length $\xi$, with $\xi$ having a weaker effect than $\Delta$.

\begin{table}
    \caption{Calculated thermal conductivity (in $\mathrm{W/m\cdot K}$) of a 24.6-nm-wide graphenelike nanoribbon calculated based on the FDTD solution to the elastic and scalar wave equations with free and fixed boundary conditions. Roughness rms value $\Delta$ and correlation length $\xi$ are measured in units of grid-cell size $h$ ($h=$0.246 nm). }
    \label{table:results}

    \begin{tabular}{l ccccc c ccccc}
    \hline \hline
        $\Delta \rightarrow$ & 1 & 2 & 3 & 4 & 5 & {}& 1 & 2 & 3 &4 & 5                        \\ \hline
        $\xi \downarrow$ & \multicolumn{5}{c}{Elastic waves, free BCs} &{}& \multicolumn{5}{c}{Scalar waves, free BCs} \\ \cline{2-6} \cline{8-12}
        3        & 1254 & 752 & 540 & 426 & 343 &{}& 2916 & 1351 & 782 & 516 & 371 \\
        6        & 1409 & 757 & 563 & 426 & 340 &{}& 3093 & 1361 & 794 & 560 & 393 \\
        9        & 1479 & 814 & 588 & 431 & 353 &{}& 3210 & 1455 & 820 & 567 & 414 \\
        12       & 1620 & 834 & 580 & 455 & 362  &{}& 3544 & 1499 & 882 & 582 & 427 \\
        15       & 1695 & 898 & 591 & 458 & 365 &{}& 3727 & 1581 & 874 & 610 & 431 \\
        \hline
        & \multicolumn{5}{c}{Elastic waves, fixed BCs} &{}& \multicolumn{5}{c}{Scalar waves, fixed BCs} \\ \cline{2-6} \cline{8-12}
        3        & 2309 & 1172 & 754 & 529 & 396 &{}& 3457 & 1411 & 842 & 564 & 417 \\
        6        & 2697 & 1267 & 750 & 573 & 399 &{}& 3530 & 1520 & 853 & 541 & 433 \\
        9        & 2934 & 1255 & 812 & 556 & 423 &{}& 4289 & 1673 & 917 & 579 & 434 \\
        12       & 3272 & 1382 & 810 & 581 & 424 &{}& 4063 & 1746 & 955 & 600 & 434 \\
        15       & 3287 & 1531 & 867 & 585 & 450 &{}& 4535 & 1812 & 1020 & 645 & 460\\
        \hline\hline
    \end{tabular}

\end{table}

\begin{figure}
 \centering
 \includegraphics[width=3 in]{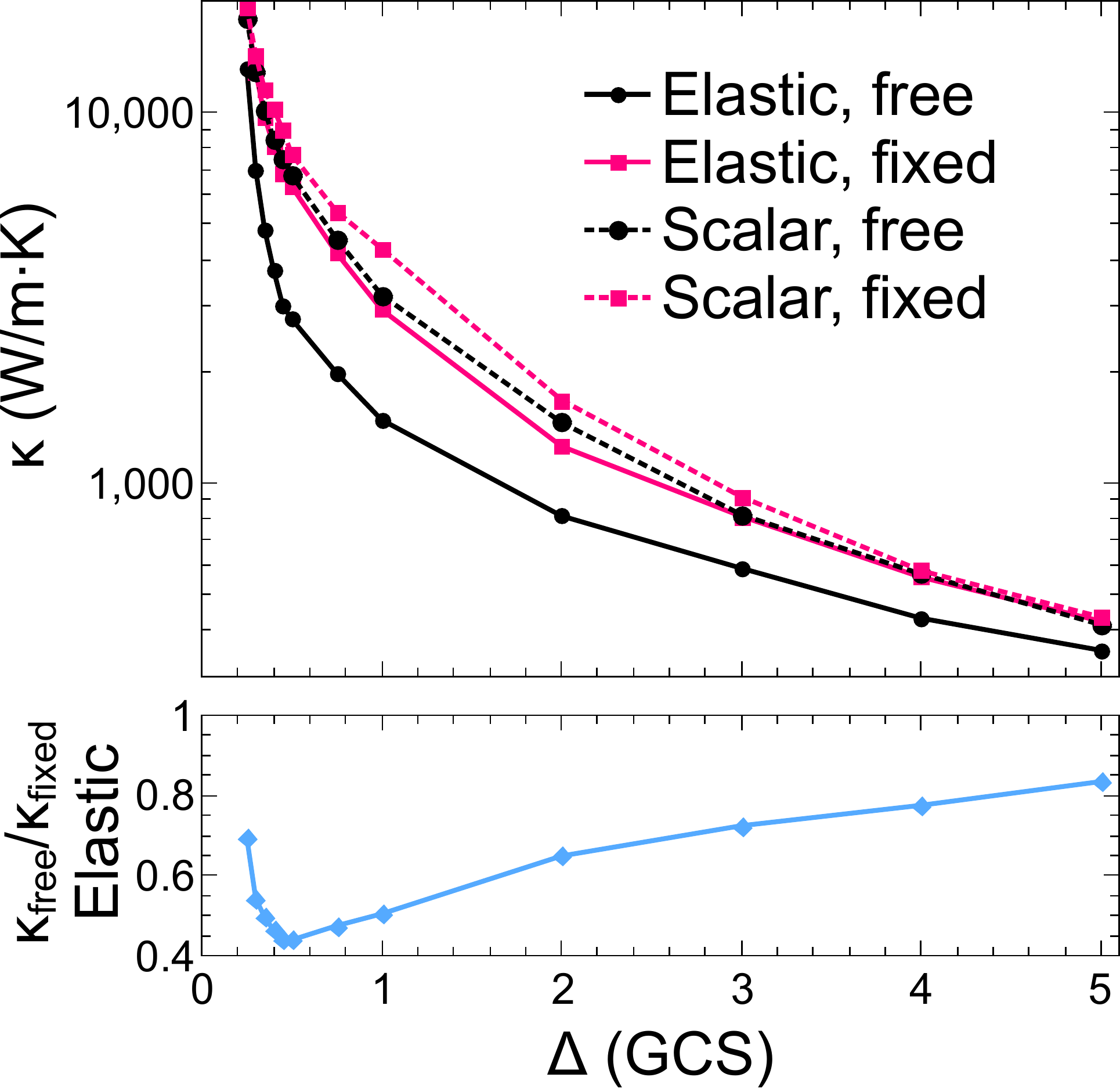}
 \caption{Top: Thermal conductivity $\kappa$ of a nanoribbon as a function of rms roughness $\Delta$ (given in the units of $h$, the grid-cell size), as obtained based on the FDTD solution to elastic (solid lines) and scalar (dashed lines) wave equations with free (black circles) and fixed (red squares) boundary conditions. The correlation length $\xi$ is fixed at $9h$ and the nanoribbon width is $100h$ (grid-cell size is $h=$0.246 nm). Bottom: Ratio of thermal conductivities resulting from free and fixed boundary conditions with the elastic wave equation. The thermal  conductivities appear to converge in the limits of no surface disorder and very high surface disorder.} \label{fig:fixed_clen}
\end{figure}

Because $\kappa$ depends weakly on $\xi$, we simplify the analysis by focusing on $\Delta$. Fig. \ref{fig:fixed_clen} shows $\kappa$ as a function of $\Delta$ for $\xi = 9h$. Scalar waves (with both BCs) and elastic waves with fixed surfaces have very similar results, which converge at large $\Delta$. Elastic waves with free BCs have significantly lower thermal conductivities. By switching from fixed to free BCs, we turn on all the effects of Rayleigh waves: the slow Rayleigh waves themselves (Sec. \ref{sec:Rayleigh}), bulk-to-Rayleigh-wave conversion (Sec. \ref{sec:mode_conversion}), and their susceptibility to localization (Sec. \ref{sec:VEA}). So, our thermal conductivity results show the combined effect of the existence of Rayleigh waves.

Although our method has significant limitations in modeling real materials, to test our simulation on larger structures, we also simulated 45-nm-wide nanoribbons with 0.6-nm rms roughness, similar to those measured by Bae \textit{et al.} \cite{Pop_NatCom_12}. Our calculation, which only accounts for edge roughness, yielded $\kappa \approx 500 \; \mathrm{W/m\cdot K}$; Bae \textit{et al.} measured $\kappa \approx 80 \; \mathrm{W/m\cdot K}$ at 300 K, considerably lower in part because of three-phonon scattering. Other simulation techniques have also predicted high thermal conductivities for rough graphene nanoribbons that are smaller than the approximately 25-nm-wide nanoribbons we consider here \cite{Keblinski_APL_10, Cuniberti_PRB_10, Roche_PRB_10}. For example, Evans \textit{et al.} \cite{Keblinski_APL_10} predicted a thermal conductivity of $\approx 3000-4000 \; \mathrm{W/m\cdot K}$ for a 11-nm-wide graphene nanoribbon with one lattice constant of rms edge roughness.

\subsection{Interpretation and Casimir's model}\label{sec:Casimir}

It may seem surprising that the scalar and fixed elastic results are nearly identical, but the results are in line with one of Casimir's insights \cite{Casimir_38}: for bulk modes scattering from reflective surfaces (i.e., surfaces that do not ``trap'' energy), sufficient roughness scatters everything diffusely; the surface behaves like a blackbody that absorbs all incident phonons and immediately radiates them away at random angles. The details of the diffuse scattering (fixed or free surface, mode conversion or not) are irrelevant. All that matters is that the surface reflects everything diffusely. In this way, Casimir avoided the issue of mode conversion, even though he considered elastic waves. In Fig. \ref{fig:fixed_clen}, we see similar results. When the roughness is low, there are some differences between the scalar and fixed elastic results, but the results all converge as the roughness increases and the scattering becomes totally diffuse. (When there is no roughness, all the results also converge since there is no meaningful surface scattering.)

However, Casimir's insight only holds for reflective surfaces. In contrast, free surfaces support Rayleigh waves, which can effectively trap incident energy at the surface instead of reflecting it (Secs. \ref{sec:mode_conversion},\ref{sec:energy_localization}). This limitation in Casimir's assumptions helps to explain why the free elastic results are different from the others, but not why they are are necessarily lower. At first glance, it might seem that Rayleigh waves should increase thermal conductivity; after all, Rayleigh waves are another mode to transport energy, and they also have the benefit of always traveling down the axis of the ribbon. However, Rayleigh waves have the disadvantage of being slower than either transverse or longitudinal waves, even when there is no disorder. More importantly, Rayleigh waves concentrate the energy where there is the most disorder, at the surface (Secs. \ref{sec:mode_conversion},\ref{sec:energy_localization}), so the energy concentrated at the surface likely contributes little to thermal transport. Finally, free surfaces with rough boundaries increase the localization of all modes, including bulk modes, possibly through mode conversion with Rayleigh waves (Sec. \ref{sec:VEA}).

Additionally, Casimir only considered planar surfaces, which puts a lower limit on the thermal conductivity in 3D systems. In 2D, Casimir's model predicts an infinite thermal conductivity in the absence of other scattering mechanisms \cite{Mingo_APL_11}. The failure of Casimir's model in 2D raises the question of why our model predicts a finite thermal conductivity. Indeed, our model does not inherently have a finite thermal conductivity: the calculated thermal conductivity diverges if the boundaries are flat, so the rough  boundaries (rather than finite length or discretization) are the cause of the finite thermal conductivities in our system. (Other systems where non-planar boundaries produce finite thermal conductivity include some 2D billiard channels \cite{Alonso_PRE_02}.)

We believe that several factors are at play to explain the effects of free surfaces on elastic waves. If there is no roughness, there is no meaningful surface scattering, and all the results converge (left side of Fig. \ref{fig:fixed_clen}). Adding even a little roughness causes the results to diverge from one another, as the effects of localization and bulk-to-Rayleigh-wave mode conversion take hold. These effects remain important for medium roughness (middle of Fig. \ref{fig:fixed_clen}). At high roughness, other effects of roughness, such as constricting the ribbon width in places, are evidently more important than the BCs, and a large amount of roughness will cause the thermal conductivities of free and fixed ribbons to once again converge (right side of Fig. \ref{fig:fixed_clen}).

\subsection{Absence of length effects}\label{sec:length}

\begin{figure}
 \centering
 \includegraphics[width=3 in]{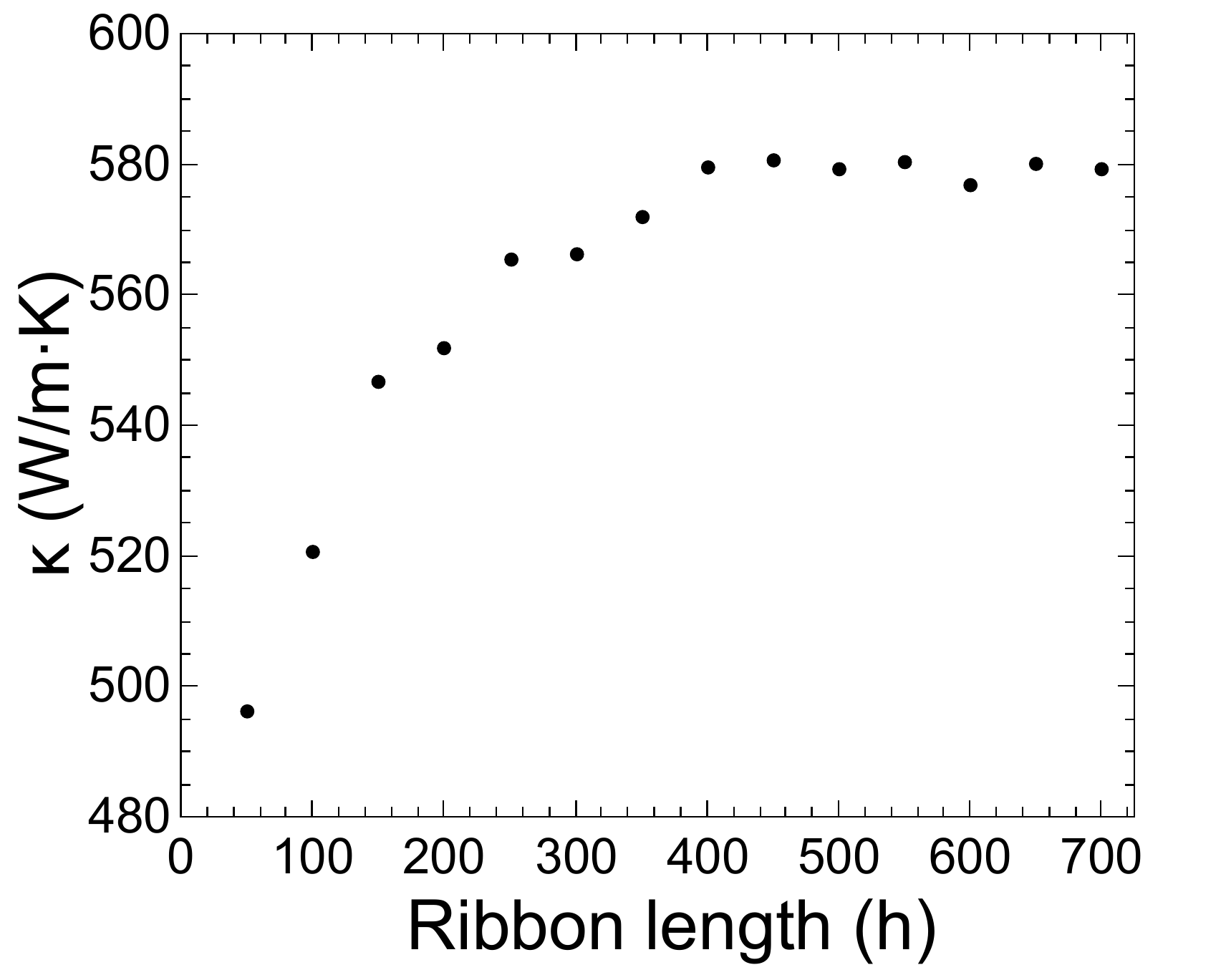}
 \caption{Thermal conductivity $\kappa$ of a $100h$-wide wire (grid-cell size $h=$0.246 nm) versus wire length. The correlation length is $9h$, and the rms roughness is $3h$. There are length effects at small lengths, but the length effects are mostly gone by a length of $300h$. Most imporantly, $\kappa$ does not appear to diverge like $\kappa \propto L^\alpha$, where $L$ is the system length and $\alpha$ is a non-zero, system-dependent constant.} \label{fig:length}
\end{figure}

As noted earlier, we chose the length of the simulation domain to be large enough to suppress length effects. See Fig. \ref{fig:length}. It may be surprising that length effects apparently disappear because it is commonly believed that, without anharmonicity, the thermal conductivity must be ill defined or diverge as the length of the structure is increased. For example, in 1D harmonic lattices (and may 1D lattices with anharmonicity), the thermal conductivity increases with system length $L$ as a power law: $\kappa \propto L^\alpha$, where $\alpha > 0$ is system dependent \cite{Li_PRL_01, Dhar_AiP_08}. In this case, it is said that the thermal conductivity is not ``finite'' since it diverges with the system length. We do not see this power-law divergence. It could be that length effects would be visible in our system for larger $L$, but a non-finite thermal conductivity is associated with a heat current autocorrelation function that has a power law decay \cite{Lepri_EPL_98}. We do not see a power law decay in the thermal conductivity either (Fig. \ref{fig:hcacf}).

First, we note that although non-finite thermal conductivities in 1D lattices have been thoroughly studied, much less is known about 2D lattices \cite{Dhar_AiP_08}. There have been reports of finite thermal conductivities in disordered, harmonic 2D lattices \cite{LYang_PRL_02}, although those results have been called into question \cite{Hu_PRL_03, Dhar_PRL_05, Dhar_AiP_08}. More importantly, our system is not simply a lattice of harmonic oscillators that are solved with high-precession integration. Our system is an approximation of the elastic wave equation in a system that is infinite in one dimension solved using the FDTD method on a finite system with periodic boundary conditions in the infinite-size direction. (FDTD studies on systems that are similar to ours have shown localization \cite{Weaver_JoSV_00} that results in an ``absence of transport'' \cite{Weaver_10}.) The elastic wave equation also allows for wave chaos. We see signatures of wave chaos in our structures by looking at the nearest-neighbor level spacings and inverse participation ratio distributions \cite{Tanner_JPA_07}; this chaotic behavior may be related to the thermal conductivity and we will discuss in a future publication.

\section{Energy localization}\label{sec:energy_localization}

\begin{figure}
 \centering
 \includegraphics[width=3 in]{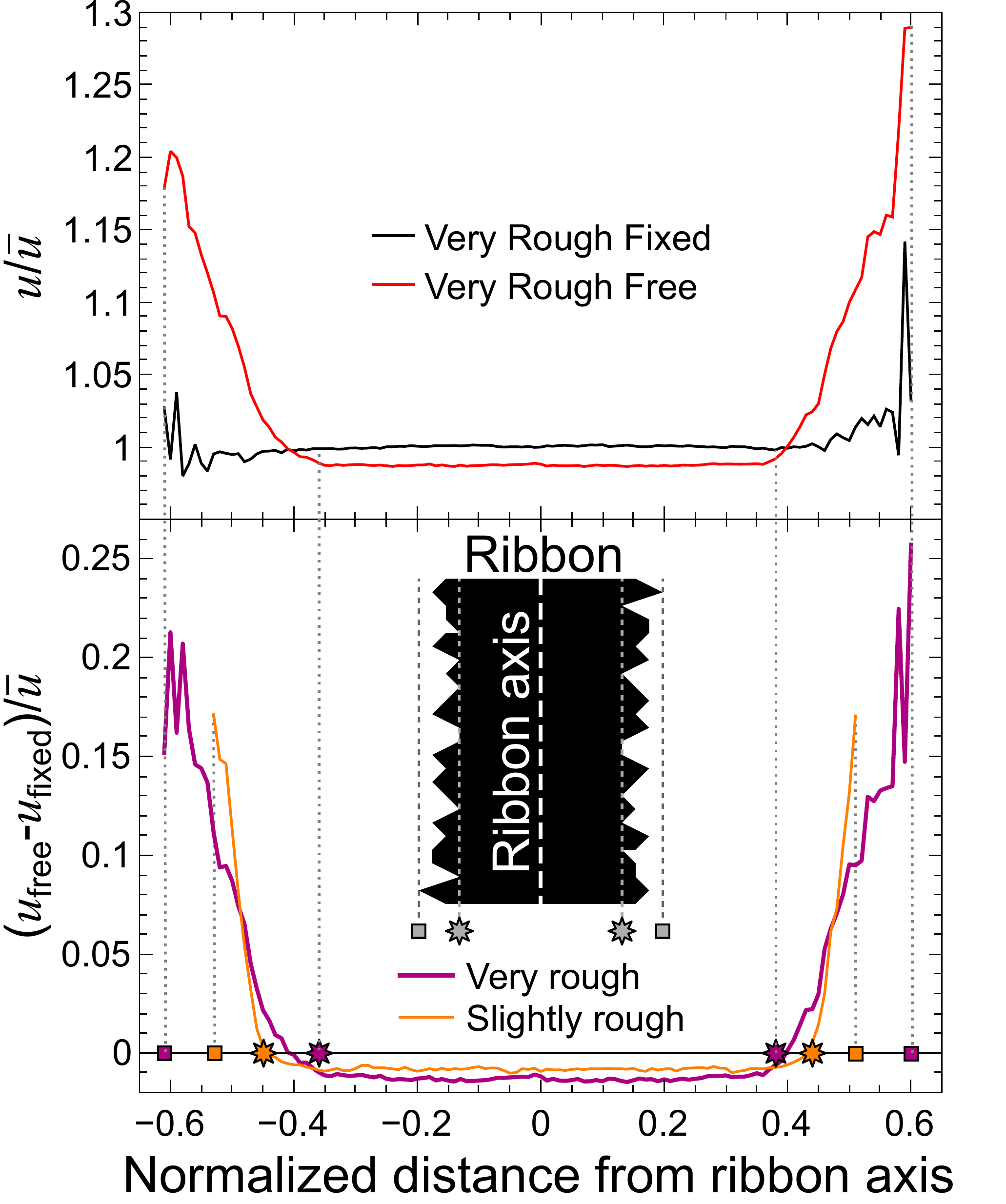}
 \caption{Spatial distributions of the energy density across the ribbon, normalized with respect to the average energy density $\bar u$. The position is given in the units of $W$, the ribbon width. Because of roughness, the distance of the edge from the nanoribbon axis varies; stars denote the minimal while squares denote the maximal distance of the edge from the axis; the star-to-star region can be considered the nanoribbon bulk. (Top) The energy densities for identical very rough ribbons ($\Delta/W=0.05$, with $\xi/W=0.09$) with free BCs (supporting Rayleigh waves) and fixed BCs  (Rayleigh waves absent) versus position with respect to the ribbon axis. The free surface siphons energy while the fixed surface has little effect on the energy profile. (Bottom) Difference in the energy densities between free and fixed BCs for very rough (purple; $\Delta/W=0.05$) and slightly rough (orange; $\Delta/W=0.02$) ribbons; $\xi/W=0.09$  for both. The free-minus-fixed energy-density profile (i.e., the difference between the curves from the top panel) represents the energy redistribution that stems from the presence of Rayleigh surface waves. It is notable that the energy gets moved away from the bulk and into the Rayleigh surface waves, an effect that becomes more pronounced with increasing roughness.}
 \label{fig:localization}
\end{figure}

The effect of Rayleigh-wave localization due to disorder is visible in Fig. \ref{fig:localization}, which depicts the difference between the energy-density profiles across the ribbon for free BCs (Rayleigh waves present) and fixed BCs (Rayleigh waves absent), normalized with the average energy density. Because we are performing an equilibrium simulation, the uneven energy distribution in the free ribbon shown in Fig. \ref{fig:localization} implies energy localization for the same reason that uneven energy distribution in a single mode implies localization. Each curve is obtained by averaging over the length; the noise in the energy density near the edges is simply because there are fewer points to average over near the edges. Distance from the axis is in the units of nanoribbon width, so the average width goes from -1/2 to 1/2. Stars denote the minimal and squares the maximal distance between an edge and ribbon axis; in other words, the star-to-star distance denotes the minimum width (we can consider this the bulk region), while the square-to-square distance is the maximal ribbon width.

As the edge roughness increases on free wires, the energy density in the center of the ribbon decreases, and the energy density near the edges increases. However, increasing roughness on fixed wires has little impact on the energy-density profile. This is surely the handiwork of the localized modes that are associated with free BCs (Sec. \ref{sec:VEA}). The increased roughness combined with free BCs must be creating more edge modes (or hybridize bulk and edge modes) with the effect that energy from the center of the wire is shifted to the edges. The edges are the most disordered part of the system, so energy stored at the edges likely contributes little to transport. This could be one way that increasing the roughness causes lower thermal conductivity.

\section{Implications for single-scalar-wave models and Phonon Monte Carlo}\label{sec:implications}

Our results have implications for methods and models used to simplify phonon--surface scattering. Simplified phonon--surface scattering models are particularly important for many modern nanostructures, which are often too large to simulate with first-principles methods \cite{Cahill_APR_14}.

Many phonon--surface scattering models use the phenomenological concept of a specularity parameter \cite{Ziman_PRS_53, Ziman_PRS_55, Ziman_Book, Soffer_JAP_67, Robinson_PRB_94, Maznev_PRB_15}, the probability a wave will scatter specularity from a rough surface. Existing specularity parameter models are tied to the scalar wave model, which does not support mode conversion. Indeed, Ziman adapted the concept of a specularity parameter from electromagnetic waves, which do not undergo meaningful mode conversion \cite{Ziman_PRS_55}. Our results underscore the importance of the ongoing work to extend the specularity parameter model to account for mode conversion \cite{Ogilvy_NDTI_86} and Rayleigh waves \cite{Maznev_PRB_15}. However, even with improvements, the specularity parameter concept only makes sense in the limit of weak roughness; all specularity parameter models have a ``Casimir limit'' when the specularity parameter is zero and all scattering is diffuse \cite{Knezevic_APL_15}, yet in many three-dimensional nanostructures thermal conductivities far below that limit have been measured \cite{Yang_Nat_08, Yang_NLet_12, Bourgeois_APL_13}. It is also not clear that an improved specularity parameter can result in finite thermal conductivity in two dimensional systems in the absence of other scattering mechanisms.

Another tool used to study phonon transport in relatively large nanostructures is phonon Monte Carlo (PMC), where a large ensemble of phonons are treated as point particles that drift and scatter \cite{Peterson_JHT_94, Lacroix_PRB_05, Baillis_JHT_08, Mazumder_JHT_10, Knezevic_PRB_12, Knezevic_APL_15}. Current PMC simulations do not allow for mode conversion or Rayleigh waves. Instead, PMC simulations either have phonons scatter specularly at the surface \cite{Baillis_JHT_08, Knezevic_PRB_12, Bourgeois_APL_13, Knezevic_APL_15} or have a specularity parameter with the reflected phonon of the same mode as the incident phonon \cite{Mazumder_JHT_10, Knezevic_APL_15}. Monte Carlo simulations for chaotic ray-splitting billiards are similar to PMC simulations, and the chaotic ray-splitting billiard simulations have been extended to allow for mode conversion between bulk modes \cite{Antonsen_PRA_92}. The technique could be extended to PMC and, in principle, could support Rayleigh waves if the scattering amplitudes between bulk and Rayleigh waves are known.

Finally, let us consider what our results mean for the many phonon--surface scattering models based on scalar waves \cite{Lambert_PRB_99, Cross_PRB_01, Lifshitz_PRB_01, Yung_PRB_01, Vasilopoulos_PRB_01, Geller_PRB_04, Chang-Long_CPL_06, KChen_PLA_06, KChen_JoPD_07, Xie_JoPD_07, Zhao_JoPD_07, JGong_PRB_09}. While the scalar wave equation significantly overestimates $\kappa$, it does produce qualitatively similar results to the elastic wave equation: $\kappa$ decreases with increasing $\Delta$ and with decreasing $\xi$. Problems arise when accurate predictions are needed for known surface roughness. This issue was seen in the work of Santamore and Cross, who studied a system first using scalar \cite{Cross_PRB_01} and then elastic wave equations \cite{Cross_PRL_01}. While both models could fit experimental data, they required significantly different surface roughness parameters. (The actual surface roughness features were unknown.) For a surface with a known roughness profile, an elastic-wave model would be more accurate than a scalar-wave model.

\section{Conclusion}\label{sec:discussion}

In this paper, we introduce the novel technique of calculating thermal conductivity by coupling a finite-difference-time-domain solution of the elastic wave equation with the Green-Kubo formula (Sec. \ref{sec:method}). This technique can handle the large structures needed to investigate highly disordered surfaces. We apply this technique to 2D, but the method can easily be extended to 3D.

By including free surfaces, we see that not only do we introduce localized surface modes (such as propagating Rayleigh waves and non-propagating wag modes), but low-frequency bulk modes also become significantly more localized (Sec. \ref{sec:VEA}). This effect is not seen with fixed surfaces, which indicates that surface disorder alone does not cause the significant increase in localization. Instead, the increased localization is caused by a combination of surface disorder and boundary conditions, likely related to mode conversion between bulk and Rayleigh waves.

This increased localization is coincident with a buildup of energy at free surfaces but not at fixed surfaces (Sec. \ref{sec:energy_localization}). This buildup reduces the energy density in the center of the ribbon. Since the edges are the most disordered part of the system, it stands to reason that the buildup of energy at the edges plays a role in reducing the thermal conductivity.

Indeed, we see that increased disorder does reduce the thermal conductivity, and this effect is more pronounced in ribbons with free boundary conditions than with fixed boundary conditions (Sec. \ref{sec:results}), which underscore the importance of Rayleigh waves in thermal conductivity lowering. The results for fixed and free ribbons appear to converge in the limit of large disorder. With enough disorder, it could be that other effects, such as significant constrictions of the ribbon, become more important than the effects of boundary conditions.

We qualitatively investigate mode conversion (Sec. \ref{sec:mode_conversion}) and find that mode conversion between bulk modes due to surfaces has little effect on thermal conductivity (Sec. \ref{sec:results}), which can be explained with insights from Casimir's model (Sec. \ref{sec:Casimir}). In contrast, mode conversion between surface and bulk modes cannot be accounted for in Casimir's model.

Finally, we consider the limitations of common phonon-surface scattering models that do not include Rayleigh waves (Sec. \ref{sec:implications}), such as models that treat phonons as a single scalar wave. While such models can qualitatively produce similar results to models based on elastic waves, a significant gap exists between the scalar wave and elastic wave predictions. Unlike single scalar waves, elastic waves are the long-wave-length limit of real phonons and should result in more accurate predictions.

We have shown that free surfaces (and the associated Rayleigh waves) play an important role in phonon-surface scattering and phonon localization, yet they are not included in most phonon--surface scattering models. However, both surface and internal scattering mechanisms (phonon-phonon, mass-difference, etc.) play important roles in nanoscale thermal transport. Additional work is needed to develop models that include internal scattering mechanisms, incorporate the effects of free surfaces, and are able to simulate large nanostructures.

\begin{acknowledgments}
The authors thank G. P. Tsoflias and C. Zeng for helpful discussions on the energy stability of the FDTD algorithm. This work was supported by the U.S. Department of Energy (Office of Basic Energy Sciences, Division of Materials Sciences and Engineering, Physical Behavior of Materials Program) under Award DE-SC0008712. The work was performed using the compute resources and assistance of the UW-Madison Center for High Throughput Computing (CHTC) in the Department of Computer Sciences.
\end{acknowledgments}

\end{document}